\documentclass[a4paper,twocolumn,10pt,accepted=2024-09-14]{quantumarticle}
\pdfoutput=1
\usepackage[utf8]{inputenc}
\usepackage[english]{babel}
\usepackage[T1]{fontenc}
\usepackage{amsmath}
\usepackage{hyperref}

\usepackage{tikz}
\usepackage{lipsum}

\usepackage{graphicx}
\usepackage{indentfirst}
\usepackage{physics}
\usepackage{braket}
\usepackage{float}
\usepackage{amsmath}
\usepackage{amsthm}
\usepackage{wasysym}
\usepackage{mathrsfs}
\usepackage{bm}
\usepackage{epstopdf}
\usepackage{mathtools}
\usepackage{CJK}
\usepackage{esint}
\usepackage{color}
\usepackage{subfigure}
\usepackage{amsfonts}
\usepackage{footmisc}
\usepackage{scrextend}
\usepackage{multirow}
\usepackage{amssymb}
\usepackage{multirow}
\usepackage{diagbox}

\usepackage[numbers,sort&compress]{natbib}
\usepackage{url}
\usepackage{bm}

\usepackage{xcolor}
\definecolor{darkblue}{rgb}{0,0,0.5}
\hypersetup{
colorlinks=true,
linkcolor=black,
filecolor=blue,
citecolor=darkblue,  
urlcolor=black,
}
\newtheorem{theorem}{Theorem}

\newtheorem{definition}{Definition}

\newcommand{\bx}{\boldsymbol x}

\newcommand{\bOmega}{\boldsymbol \Omega}

\newcommand{\bV}{\boldsymbol{V}}
\newcommand{\bI}{\boldsymbol I}
\newcommand{\bZ}{\boldsymbol Z}
\newcommand{\bA}{\boldsymbol A}
\newcommand{\bM}{\boldsymbol M}
\newcommand{\bN}{\boldsymbol N}
\newcommand{\bd}{\boldsymbol d}
\newcommand{\bR}{{\boldsymbol {R}}}

\newcommand{\calC}{{\cal{C}}}

\newcommand{\calL}{{{\cal{L}}}}

\newcommand{\Sp}{{\rm{Sp}}}
\newcommand{\be}{{\boldsymbol{e}}}
\newcommand{\bS}{{\boldsymbol {S}}}
\newcommand{\bB}{\boldsymbol {B}}
\newcommand{\hrho}{\hat{\rho}}
\newcommand{\Real}{\mathbb{R}}
\newcommand{\Integer}{\mathbb{Z}}

\newcommand*\diff{\mathop{}\!\mathrm{d}}

\newcommand{\QZ}[1]{{{\textcolor{black}{#1}}}}

\begin{document}

\title{Safeguarding Oscillators and Qudits with Distributed Two-Mode Squeezing}

\author{Anthony J. Brady}\email{ajbrad4123@gmail.com}
\affiliation{
Ming Hsieh Department of Electrical and Computer Engineering, University of Southern California, Los
Angeles, California 90089, USA
}

\author{Jing Wu}
\email{jingwu@arizona.edu}
\affiliation{James C. Wyant College of Optical Sciences, University of Arizona, Tucson, AZ 85721, USA}

\author{Quntao Zhuang}
\email{qzhuang@usc.edu}
\affiliation{
Ming Hsieh Department of Electrical and Computer Engineering, University of Southern California, Los
Angeles, California 90089, USA
}
\affiliation{
Department of Physics and Astronomy, University of Southern California, Los
Angeles, California 90089, USA
}


\begin{abstract}
    Recent advancements in multi-mode Gottesman-Kitaev-Preskill (GKP) codes have shown great promise in enhancing the protection of both discrete and analog quantum information. This broadened range of protection brings opportunities beyond quantum computing to benefit quantum sensing by safeguarding squeezing---the essential resource in many quantum metrology protocols. 
    However, the potential for quantum sensing to benefit quantum error correction has been less explored. In this work, we provide a unique example where techniques from quantum sensing can be applied to improve multi-mode GKP codes. 
    Inspired by distributed quantum sensing, we propose the distributed two-mode squeezing (dtms) GKP codes that offer benefits in error correction with minimal active encoding operations. Indeed, the proposed codes rely on a \textit{single} (active) two-mode squeezing element and an array of beamsplitters that effectively distributes continuous-variable correlations to many GKP ancillae, similar to continuous-variable distributed quantum sensing. Despite this simple construction, the code distance achievable with dtms-GKP qubit codes is comparable to previous results obtained through brute-force numerical search [PRX Quantum 4, 040334 (2023)]. Moreover, these codes enable analog noise suppression beyond that of the best-known two-mode codes [Phys. Rev. Lett. 125, 080503 (2020)] without requiring an additional squeezer. We also provide a simple two-stage decoder for the proposed codes, which appears near-optimal for the case of two modes and permits analytical evaluation.
    
\end{abstract}


\maketitle



\section{Introduction}
Quantum error correction is essential for robust quantum information processing, as quantum effects such as entanglement and coherence are otherwise susceptible to ubiquitous environmental noise. Bosonic codes~\cite{albert2018pra,terhal2020towards,joshi2021quantum,albert2022bosonic,brady2023GKPrvw} are among the first codes that have led to experiments with extended coherent storage of quantum information beyond break-even~\cite{ofek2016extending,sivak2022breakeven}. Exploiting the infinite dimensional Hilbert space of oscillators, bosonic codes are hardware efficient and promise hope to fault-tolerant quantum computing. In particular, recent progress in codes based on Gottesman-Kitaev-Preskill (GKP) states provides a versatile platform that safeguards discrete-variable information by encoding qubits into oscillators~\cite{gkp2001,harrington2001rates,noh2018GKPcapacity} and continuous-variable information through oscillators-to-oscillators (O2O) encodings~\cite{noh2020o2o,wu2022continuous,wu2023optimal}, as summarized in a recent review~\cite{brady2023GKPrvw}.

The broadened scope of quantum information protection not only supports fault-tolerant quantum computing but also extends the utility of GKP codes to enhance various quantum sensing protocols by protecting metrological resources such as squeezing and entanglement~\cite{zhuang2020GKP_dqs,zhou2022enhancing}. On the other hand, it is an open question how quantum sensing techniques can help with quantum error correction, considering that measurement in quantum sensing is often destructive since only classical information, such as unknown parameters, is of interest. We approach this question by considering practical resource constraints involved in bosonic codes. Indeed, for bosonic quantum sensing, where infinite dimensional Hilbert space is concerned, advantages are always analyzed with constraints so that the problem becomes regularized.


The major resources required to engineer multi-mode GKP codes~\cite{conway2013sphere,baptiste2022multiGKP,eisert2022lattice,brady2023GKPrvw,lin2023closest,xu2022qubit,conrad2023ntru} involve the capability of generating standard single-mode square-lattice GKP states and applying inline quantum operations on them, a combination that ensures universal engineering of such codes. While the engineering of a single-mode square lattice GKP state is challenging, it can be done {\it offline} and recent experiments have shown promise in microwaves~\cite{devoret2020GKPnature,eickbusch2022fast,sivak2022breakeven}, trapped ions~\cite{home2019GKPnature,home2022QECgkp}, and optics domains~\cite{konno2023propagatingGKP}. Moreover, single-mode square lattice GKP states have been chosen in various works~\cite{baragiola2019GKPuniversal, bourassa2021blueprint,noh2020o2o,xu2022qubit,wu2023optimal} as the standardized non-Gaussian resource~\cite{zhuang2018resource,takagi2018convex} that is necessary for universal, fault-tolerant quantum information processing~\cite{eisert2002nogodistillation,fiuravsek2002gaussian,niset2009nogo}. On the other hand, inline quantum operations---specifically active components such as single-mode squeezing and two-mode squeezing operations---can be challenging to implement with high efficiency. For this reason, efforts are devoted to designing universal quantum processors with only inline passive Gaussian operations acting on off-line prepared GKP states and squeezed vacuum states~\cite{walshe2020CVteleport}.

\begin{figure*}
    \centering
    \includegraphics[width=.9\linewidth]{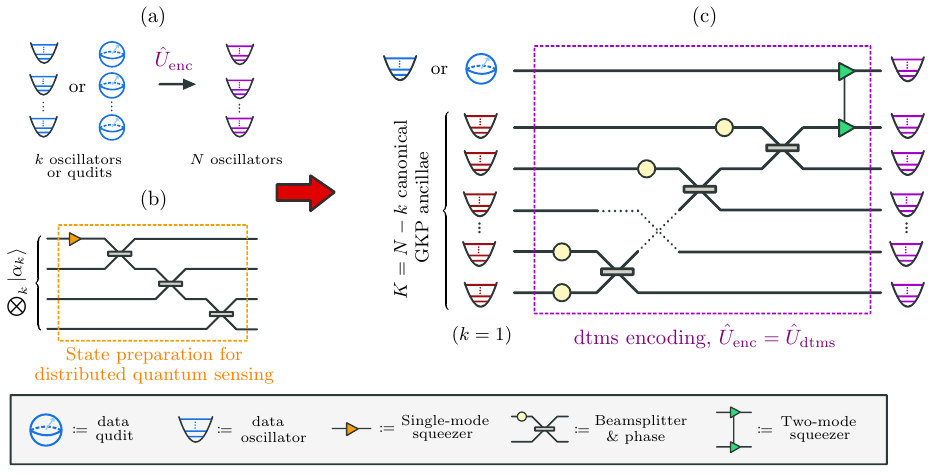}
    \caption{Combining ideas from analog and discrete quantum information. (a) Encoding oscillators or qudits into many oscillators. $k$ data oscillators (e.g., multi-partite entangled state) or $k$ data qudits (e.g., GKP square qubits, $d=2$) are encoded into a system of $N>k$ oscillators via $\hat{U}_{\rm enc}$. (b) Distributed quantum sensing. A multi-partite entangled state is prepared by generating and subsequently distributing a bright squeezed beam (i.e., a squeezed coherent state) to many modes. Inputs are separable probes, either vacuum or coherent states ($\bigotimes_k\ket{\alpha_k}$). This setup offers quantum enhanced sensing of global parameters~\cite{Zhang2021dqs}, such as an average amplitude or phase. (c) Distributed two-mode squeezing (dtms) GKP codes introduced in this paper. Schematic of the dtms-GKP encoder for $k=1$, inspired by oscillator codes and distributed quantum sensing. The unitary Gaussian encoder $\hat{U}_{\rm dtms}$ comprises a single two-mode squeezer of gain $G$ and a staircase of beamsplitters with transmissivities $\eta_j$ and phases $\phi_j$. The block of $K=N-1$ oscillators consists of canonical ($d=1$) square GKP ancillae.}
    \label{fig:dtms_schematic}
\end{figure*}

In this work, we tackle the resource-efficient multi-mode GKP code design problem, focusing on minimizing the number of active inline Gaussian operations. Inspired by oscillators-to-oscillators codes~\cite{noh2020o2o} and continuous-variable distributed quantum sensing~\cite{zhuang2018dqs,Zhang2021dqs}, we propose the {\it distributed two-mode squeezing (dtms) GKP codes}---applicable to both qudits and oscillators---that employ only a single active element (a two-mode squeezer) and beamsplitters. The codes function by uniformly distributing squeezing to all modes and utilizing the resulting continuous-variable correlations to improve error correction, analogous to distributed quantum sensing. See Fig.~\ref{fig:dtms_schematic} for an illustrative schematic.


In terms of code distance, our proposed dtms-GKP qubit codes achieve comparable performance to the codes identified (via brute-force numerical search) in Ref.~\cite{lin2023closest}, while offering a significantly simplified construction. For analog error correction (i.e., O2O codes), the dtms-GKP codes provide additional enhancement beyond the quadratic error suppression found in Ref.~\cite{noh2020o2o}, without introducing additional inline squeezing elements. To illustrate with concrete examples, we identify a dtms-GKP code that encodes two qubits into three modes and surpasses square and hexagonal GKP qubit encodings. For a single qubit encoded into four modes, we find a dtms-GKP code that outperforms the conventional [[4,1,2]] qubit code~\cite{vaidman1996_42code,grassl1997_42code,leung1997approxQEC} concatenated with a local GKP code~\cite{rozpkedek2021quantum}. 
This superior performance is achieved using two beamsplitters and a single two-mode squeezing operation of approximately $7$dB. 
We also provide simple two-stage linear decoders, which appear near-optimal for the case of two modes, and evaluate performance analytically. These compact codes are particularly well-suited for applications in quantum repeaters~\cite{azuma2023RMPQuInternet,fukui2021allOptical,rozpkedek2021quantum,wu2022continuous,schmidt2023RepeaterGKPqudits,rozpedek2023allphotonic} and quantum sensor networks~\cite{zhuang2020GKP_dqs}.

\section{High-Level Overview}\label{sec:summary}

We provide a high-level overview of the key conceptual elements, main results, and contextual relevance of our paper. Technical machinery, derivations, and quantitative analyses can be found in the main body of the paper, with some details given in the appendices. First, we provide the background needed to understand the open problem that we solve. Next, we provide a detailed summary of the main results.

\subsection{Multi-mode GKP codes: Qudits-to-oscillators and oscillators-to-oscillators}
Our work concerns bosonic codes that utilize GKP states to encode finite dimensional systems (e.g., a qudit with $d$ levels) into the infinite-dimensional Hilbert space of the oscillators~\cite{gkp2001}. This encoding is achieved via lattice-like structures in phase space, resembling quadrature amplitude modulation codes in classical optical communication~\cite{girvin2023notes}. GKP codes have recently gained prominence in practical bosonic quantum information processing~\cite{terhal2020towards,joshi2021quantum,albert2022bosonic,brady2023GKPrvw}, offering a potential path to optical fault-tolerant quantum computation~\cite{baragiola2019GKPuniversal, bourassa2021blueprint,walshe2020CVteleport} and long-distance quantum communication~\cite{azuma2023RMPQuInternet,fukui2021allOptical,rozpkedek2021quantum,wu2022continuous,schmidt2023RepeaterGKPqudits,rozpedek2023allphotonic}. 

The logical computational basis for a square-grid GKP qudit, expressed in the position eigenstates of the oscillator, is
\begin{equation}\label{eq:gkp_comp_basis}
\ket{j;d}\coloneqq\sum_{n\in\mathbb{Z}}\ket{\sqrt{2\pi/d}(d n-j)}_{\hat{q}},
\end{equation}
where $j=0,1,\dots, d-1$ denotes the $d$ levels, $n$ sums over all integers $\mathbb{Z}$ in the ideal case and $\ket{\cdot}_{\hat{q}}$ denotes the position eigenstate. These discrete basis states span a $d$-dimensional subspace (code space) $\mathcal{C}$ of the infinite-dimensional bosonic Hilbert space of the oscillator. A generic GKP code word $\ket{\Psi}\in\mathcal{C}$ can be expanded in the computational basis in the usual way. In phase space, the GKP state above represents a square lattice.

Pauli $X$ and $Z$ operations on a GKP qudit correspond to linear translations in phase space,
\begin{equation}\label{eq:qudit_paulis}
    \hat{X}= \exp\left(i{\sqrt{\frac{2\pi}{d}}}\hat{p}\right) \qq{and} \hat{Z}= \exp\left(-i\sqrt{\frac{2\pi}{d}}\hat{q}\right),
\end{equation}
where $\hat{p}$ and $\hat{q}$ are the momentum and position quadrature operators.
Similar to conventional two-level systems, the generalized Pauli $X$ and $Z$ operations for $d$-level systems induce jumps and phase flips on the computational basis states, i.e.
\begin{align}
    \hat{X}\ket{j;d}&=\ket{(j+1) {\rm mod}\, d;d} \\
    \qq{and}
    \hat{Z}\ket{j;d}&=e^{i2\pi j/d}\ket{j;d}.
\end{align}

An interesting state is the canonical GKP state, $\ket{\rm GKP}\coloneqq \ket{0;1}$, with only a single state in the family as $d=1$. By itself, the canonical GKP state cannot encode quantum information. However, it is a useful ancillary resource in multi-mode GKP qudit codes and, furthermore, facilitates analog quantum quantum error correction~\cite{noh2020o2o}.

Two types of GKP codes---multi-mode qudit codes~\cite{conway2013sphere,baptiste2022multiGKP,eisert2022lattice,brady2023GKPrvw,lin2023closest,xu2022qubit,conrad2023ntru} and oscillators-to-oscillators codes~\cite{noh2020o2o,xu2022qubit,wu2023optimal}---have been recently proposed to further enhance error correction capabilities. Our work applies to both types of codes and, thus, we introduce both codes below. 

We begin with multi-mode qudit codes. The GKP state referred to in Eq.~\eqref{eq:gkp_comp_basis} corresponds to a single-mode encoding. To generate a multi-mode qudit code, we start with a collection of $k$ GKP qudits, one for each bosonic mode. The $t$-th mode encodes a qudit of $d_t$ levels. We can then employ a block of $N-k$ canonical GKP ($d=1$) ancillae to encode the $k$ qudits into $N$ modes by coupling all the subsystems through a multi-mode Gaussian unitary, $\hat{U}_{\rm enc}$, {\it viz.}
\begin{align}\label{eq:ktoN_GKP_qudits}
\ket{j_1,\cdots,j_k}_{k\to N}=\hat{U}_{\rm enc} \left[\left(\bigotimes_{t=1}^k \ket{j_t; d_t}\right)\otimes \ket{\rm GKP}^{\otimes (N-k)}\right],
\end{align}
where the subscript $k\to N$ denotes the number of modes in the encoding. See Fig.~\ref{fig:dtms_schematic}(a) for an illustration. 

Now we introduce the oscillators-to-oscillators GKP codes.
To protect an analog quantum state, such as a bright squeezed beam or a two-mode squeezed vacuum state, we follow a similar approach to the qudit case. However, in this instance, the $k$ data modes harbor a multi-mode continuous-variable state $\ket{\varphi}_k$, e.g., a multi-partite entangled state that is beneficial for quantum sensing~\cite{zhuang2018dqs}. It has been known for quite some time that Gaussian states and Gaussian operations (i.e., squeezers, beamsplitters, and phase shifts) alone cannot protect an arbitrary oscillator state~\cite{eisert2002nogodistillation,fiuravsek2002gaussian,niset2009nogo}. However, the utilization of canonical GKP ancillae in the code, owing to their non-Gaussian character, bypasses these no-go limitations~\cite{noh2020o2o}. In Ref.~\cite{wu2023optimal}, a general multi-mode encoding was considered as
\begin{align}
\ket{\varphi}_{k\to N}=\hat{U}_{\rm enc} \left(\ket{\varphi}_k \otimes  \ket{\rm GKP}^{\otimes (N-k)}\right).
\end{align}

In both multi-mode qudit codes and oscillators-to-oscillators codes, the code design involves choosing the number of modes $N$ to encode $k$ data modes, and more importantly the choice of a unitary Gaussian encoder $\hat{U}_{\rm enc}$. 
To achieve the best performance, generic code optimizations were explored in Ref.~\cite{wu2023optimal} for O2O codes and Ref.~\cite{lin2023closest} for multi-mode qubit codes. However, the results heavily rely on numerical optimization and the resource constraints are not taken into consideration in these works, leading to code designs that are potentially challenging to realize.
Indeed, the unitary Gaussian encoder $\hat{U}_{\rm enc}$ can be generic, meaning that it could, in principle, require many inline active elements sandwiched between multi-port interferometers. 
Inline active operations, such as single- and two-mode squeezing, are in general challenging to implement. In the optical domain, such inline squeezing elements require a high nonlinearity while maintaining a high quantum efficiency. In the microwave domain, inline squeezing will require the precise control of a sequence of quantum gates~\cite{eickbusch2022fast} in a cavity-QED system. On the one hand, we want to reduce the number of squeezing elements as much as possible. On the other hand, it is known that some amount of squeezing is necessary for good code performance using unitary Gaussian encoders~\cite{hanggli2022NoThresholdGKP,eisert2022lattice,wu2023optimal}. Thus, a trade-off exists to limit the amount of squeezing (or number of squeezers) without sacrificing too much in code performance. In light of these practical considerations, in this work, we consider minimal code constructions of $\hat{U}_{\rm enc}$ to simplify and facilitate experimental implementation.

\subsection{Main contributions}
We introduce a family of codes---applicable to both qudits and oscillators---that employ only a single active element (a two-mode squeezer) and a multi-port interferometer. The code operates by first establishing correlations between the noises of the $k$ data modes ($k$ GKP qudits or a $k$-mode continuous-variable oscillator state) and the $N-k$ canonical GKP ancillae through two-mode squeezing interactions. Subsequently, the correlated noises are uniformly distributed among all ancillary modes through the interferometer.\footnote{This technically represents a decoding perspective (associated with $\hat{U}_{\rm dec}=\hat{U}^{-1}_{\rm enc}$) following the noise process, which is valuable for illustrative purposes.} For the $k=1$ case, the unitary encoder is
\begin{equation}\label{eq:unitary_dtms}
\hat{U}_{\rm enc}^{(k=1)}=  \left(\hat{U}_{\bm S_G}\otimes \hat{I}_{N-2}\right)\left(\hat{I}_1\otimes \hat{U}_{\bm B}\right)\eqqcolon \hat{U}_{\rm dtms},
\end{equation}
where $\hat{U}_{\bm B}$ represents the multi-port interferometer, $\hat{I}_{M}$ is the identity operation on $M$ modes, and $\hat{U}_{\bm S_G}$ represents two-mode squeezing parameterized by the gain $G$. The inverse process is used in decoding, $\hat{U}_{\rm dec}=\hat{U}_{\rm enc}^{-1}$. See Fig.~\ref{fig:dtms_schematic}{\color{blue} (c)} for a high-level circuit schematic. \QZ{We note that the decoding operation has the structure of first applying a two-mode squeezing and then applying a beamsplitter to distribute the two-mode squeezing to multiple modes. Such a structure resembles that in a distributed quantum sensing protocol~\cite{zhuang2018dqs} (see Fig.~\ref{fig:dtms_schematic}(b)), except that, in the original sensing protocol, only single-mode squeezing is considered.}

Intuitively, the success of the code originates from the strategic splitting and distribution of the locally amplified noises (amplified by the two-mode squeezing device) across the larger block of ancillary modes. The distribution process allows higher levels of squeezing and, consequently, enhanced performance with increasing number of modes. Due to the physical interpretation of the process, we call these codes \textit{distributed two-mode squeezing (dtms) GKP codes}, denoting the encoder as $\hat{U}_{\rm dtms}\coloneqq\hat{U}_{\rm enc}$, as indicated in Eq.~\eqref{eq:unitary_dtms}. We will often call the codes dtms codes for simplicity, as it is clear that all instances refer to GKP codes.

\begin{table}[t]
\renewcommand{\arraystretch}{1.25}

    \centering
    \begin{tabular}{c | c | c}
    \hline\hline
        Code & \# of modes & \# of act. compo.   \\
        \hline
        dtms & $N$ & 1  \\
        ``Generic'' GKP & $N$ & $\order{N}$ \\
        Repetition~\cite{lin2023closest} & $N$ & $\order{N}$ \\ 
        Tesseract~\cite{baptiste2022multiGKP} & 2& 1  \\
         Square-$[[4,2,2]]$ & 4 & 5 \\
         \hline\hline
    \end{tabular}
    \caption{\QZ{Assorted codes vs. number of active gates (e.g., squeezers, SUM gates) in the encoder required to construct the code starting from square-GKP states. For example, repetition-type codes demand $\order{N}$ SUM gates (analog CNOTs~\cite{gkp2001}) for encoding. }}
    \label{tab:act comp}
\end{table}


These dtms-GKP codes, despite their apparent simplicity, exhibit impressive performance, therefore bypassing the need for many active elements in creating effective multi-mode bosonic codes; \QZ{refer to Table~\ref{tab:act comp} and below for further discussion}. For instance, when encoding a single qubit into four modes, we discover a four-mode dtms-GKP code that outperforms a conventional [[4,1,2]] GKP-qubit code. Additionally, this dtms-GKP code requires only $\lesssim7$dB of squeezing. Another notable example is the simplest dtms qubit code involving 1 data qubit coupled to 1 canonical GKP state via two-mode squeezing ($N=2$ and $k=1$). Interestingly, in this scenario, we find that the resulting dtms qubit code is akin to the recently discovered Tesseract qubit code~\cite{baptiste2022multiGKP}. When encoding two logical qubits with a single two-mode squeezing operation, we find a three-mode code ($N=3$ and $k=2$) that is better than the square code and a four-mode code ($N=4$ and $k=2$) that matches the standard $[[4,2,2]]$ code in terms of code distance. Notably, our dtms code ($N=4,k=2$) only requires a single active Gaussian gate, while a direct concatenation with a $[[4,2,2]]$ qubit code will require five active Gaussian gates \QZ{ through a standard quantum circuit construction using CNOT gates~\cite{roffe2019quantum}. Note that an analog version of the CNOT is the so-called SUM gate~\cite{gkp2001}, which is an active two-mode (Gaussian) gate~\cite{tzitrin2020ProgressGKP,Liu2024HybridProcessors}. More broadly, any ``generic'' GKP stabilizer code on $N$ modes can, in principle, be constructed by applying some $N$-mode Gaussian unitary on $N$ square GKP states (see Appendix~\ref{app:theory_gkp} for the theory of GKP codes). By a standard analog gate decomposition~\cite{braunstein05}, such a construction requires at least $N$ single-mode squeezers. Contrariwise, the dtms codes proposed herein calls for a \textit{single} two-mode squeezer for $N$ modes. A specific approach of constructing an $N$-mode code is to adopt concatenation to obtain repetition codes, as detailed in Ref~\cite{lin2023closest}. There the same $\sim N$ number of active components such as the SUM gate is required.} 

We also introduce a simple linear decoder that facilitates analytical calculations for oscillators-to-oscillators codes as well as code distances and error rates for dtms qudit codes. Notably, the proposed decoder demonstrates near-optimal performance for the Tesseract-like qubit code mentioned previously. In terms of code distance, our proposed qubit codes achieve comparable performance to the codes identified (through a generic numerical search) in Ref.~\cite{lin2023closest}, while offering a significantly simplified construction using basic elements (a single squeezer and beamsplitters). We illustrate this with numerical calculations of code distance for $N\leq 5$ modes. 

Concerning oscillators-to-oscillators codes and assuming additive noise with standard deviation $\sigma$, our codes enable analog noise suppression \textit{viz.} ${\sigma^2\rightarrow\sigma^4/(N-1)}$. This indicates noise reduction beyond that of Ref.~\cite{noh2020o2o} without introducing additional inline squeezing elements.

The dtms codes introduced here are inspired by discoveries in continuous-variable quantum information processing, such as oscillators-to-oscillators codes~\cite{noh2020o2o,wu2021continuous,wu2023optimal}, which utilize two-mode squeezing to safeguard analog oscillator states, and the distributed quantum sensing paradigm~\cite{zhuang2018dqs,Zhang2021dqs}, which demonstrates the utility of a single distributed continuous-variable resource (e.g., squeezing) for enhanced quantum sensing (See Fig.~\ref{fig:dtms_schematic}). Two-mode squeezing, in particular, plays a crucial role in other continuous-variable quantum information processing tasks, such as entanglement-assisted classical communication~\cite{bennett1999entanglement,shi2020practical,hao2021entanglement}, continuous-variable teleportation~\cite{pirandola2006quantum} and quantum-dense metrology~\cite{steinlechner2013quantum}. Our findings further elucidate the value of two-mode squeezing and distributed quantum resources in both discrete and analog quantum error correction.

On a practical note, our analyses have broad applicability to GKP codes in itinerant-mode setups, such as optical quantum information processors. Notable examples include fault-tolerant quantum computing architectures based on optical GKP states~\cite{larsen2021,bourassa2021blueprint,tzitrin2021staticFTcomp, bohan2020CVcomputing} and quantum repeaters~\cite{azuma2023RMPQuInternet,fukui2021allOptical,rozpkedek2021quantum,wu2022continuous,schmidt2023RepeaterGKPqudits,rozpedek2023allphotonic}.


\section{Preliminaries}\label{sec:prelims}

In this section, we establish notation and standard quantities used throughout the paper. We work in natural physical units ($\hbar=1$). We use boldface to indicate finite-dimensional vectors (e.g., $\bm\xi\in\mathbb{R}^{2N}$) and matrices, and quantum operators are denoted by hat on the top. Vectors are in column form.

A system of $N$ quantum harmonic oscillators (modes) can be described by the position and momentum operators $\hat{\bx}\coloneqq (\hat{q}_1, \hat{p}_1,...,\hat{q}_N, \hat{p}_N)^\top$, which have a continuous spectrum over the phase space $\Real^{2N}$.  The canonical commutation relations between the $2N$ position and momentum operators can be written in compact notation as
\begin{equation}
[\hat{\bx},\hat{\bx}^\top]=i\bOmega\hat{\mathbb{I}} \qq{with}
\bOmega=\bigoplus_{i=1}^N
\begin{pmatrix}
    0 & 1\\
    -1 & 0
\end{pmatrix}.\label{eq:CCR}
\end{equation}

A Gaussian unitary operation~\cite{weedbrook2012rmp} $\hat{U}_{\bS,\bd}$ acts on the quantum bosonic system as 
\begin{align}\label{eq:gaussian_unitary}
    \hat{U}_{\bS,\bd}^\dagger \hat{\bx} \hat{U}_{\bS,\bd} = \bS \hat{\bx}+\bd,
\end{align}
where $\bm S$ is a $2N\times2N$ real matrix and $\bm d\in\mathbb{R}^{2N}$. The transformation must preserve the commutation relations [Eq.~\eqref{eq:CCR}], implying that $\bS \bOmega\bS^\top = \bOmega$. Any matrix $\bS$ satisfying this condition is said to be a symplectic matrix and represents an element of the ($2N$-dimensional, real) symplectic group $\Sp(2N,\Real)$. 

A special class of Gaussian transformations are (symplectic) orthogonal transformations, $\bB \in \Sp(2N,\Real) \cap O(2N,\Real)$, where $O(2N,\Real)$ is the orthogonal group. Practically, any symplectic orthogonal transformation $\bm B$ corresponds to a multi-port interferometer. They are regarded as passive Gaussian unitaries as they preserve the mean occupation number.


The two-mode symplectic operations (corresponding to Gaussian unitaries) relevant to our paper are the canonical two-mode squeezing (TMS) operation, $\bm S_G$, and a variable beamsplitter, $\bm B_{m,n}$. The TMS operation can be represented by a symplectic matrix 
\begin{equation}
    \bm S_G= 
    \begin{pmatrix}
        \sqrt{G}\bm I_2 & \sqrt{G-1}\bm Z \\
        \sqrt{G-1}\bm Z & \sqrt{G}\bm I_2
    \end{pmatrix},
    \label{eq:S_TMS}
\end{equation}
where $\bm Z$ is the Pauli Z matrix and $\bm I_k$ is the identity matrix of dimension $k$. For the variable beamsplitter, let indices $m,n$ represent the interaction between the $m$-th and $n$-th mode. \QZ{Related to two-mode squeezing, a single-mode squeezing can be represented by a $2\times 2$ symplectic matrix
\begin{equation}
\bm S^{(1)}_G=\begin{pmatrix}
        \sqrt{G} & 0 \\
        0 & 1/\sqrt{G}
    \end{pmatrix}.
     \label{eq:S_SMS}
\end{equation}}

The matrix $\bm B_{m,n}$ of a variable beamsplitter, with transmissivity $\cos^2\theta$ and phase $\phi$, is given by
\begin{equation}
    \bB_{m,n}(\theta,\phi)=
    \begin{pmatrix}
        \cos{\theta}\bR(\phi) & -\sin{\theta}\bR(\phi) \\
        \sin{\theta}\bI_2 & \cos{\theta}\bI_2
    \end{pmatrix}.
\end{equation}
where $\bR(\phi)$ is a single-mode phase rotation represented by
\begin{equation}\label{eq:phase}
    \bR(\phi) = 
    \begin{pmatrix}
        \cos{\phi} & -\sin{\phi} \\
        \sin{\phi} & \cos{\phi}
    \end{pmatrix}.
\end{equation}
We let ${\bm B_{50:50}\coloneqq \bm B(\pi/4,0)}$ denote a 50:50 beamsplitter. Operationally, a variable beamsplitter can be constructed from a Mach-Zehnder interferometer utilizing 50:50 beamsplitters and two variable phase shifts ($\theta,\phi$). Moreover, a general multi-port interferometer, represented by an $N$-mode orthogonal transformation, $\bm B$, can be constructed from an array of variable beamsplitters~\cite{reck94}.

\QZ{
For completeness, we introduce the SUM gate, though we do not explicitly make use of such in this paper. The unitary SUM gate is given as $\hat{U}_{\textbf{SUM}}=e^{-i\hat{q}_1\hat{p}_2}$. A symplectic matrix representation is,
\begin{equation}\label{eq:sum_gate}
    \textbf{SUM}=
    \begin{pmatrix}
        \bm I_2 & -\bm\Pi_p \\ 
        \bm\Pi_q & \bm I_2
    \end{pmatrix},
\end{equation}
where $\bm\Pi_q={\rm diag}(1,0)$ and $\bm\Pi_p={\rm diag}(0,1)$ are projections onto the $q$ and $p$ subspaces, respectively. The SUM gate is analog to the CNOT gate for GKP states~\cite{gkp2001,tzitrin2020ProgressGKP} and is useful for, e.g., ancilla-assisted syndrome measurements and handling discrete variable codes, such as repetition codes. Observe that the SUM gate is an \textit{active} gate since $\textbf{SUM}\cdot\textbf{SUM}^\top\neq\bm I_4$.
}

We now explain the noise model relevant to bosonic error correction---the quantum random displacement error channel for a $N$-mode bosonic system,
\begin{equation}\label{eq:angn_channel}
    \Phi_{P} (\hrho)=\int_{\mathbb{R}^{2N}} \diff^{2N}{\bm e} \; P(\bm e) \hat{D}(\bm e) \hrho \hat{D}^\dagger (\bm e) ,
\end{equation}
where $P(\bm e)$ is the probability density function (PDF) of random displacement $\bm e$. The displacement operator is 
\begin{equation}\label{eq:displacement}
    \hat{D}(\bm d)= e^{-i \bm d^\top \bOmega \hat{\bx}} 
\end{equation}
which, following Eq.~\eqref{eq:gaussian_unitary}, is a Gaussian unitary with $\hat{D}(\bd)= U_{\bm I,\bm d}$ and therefore induces displacement $\hat{\bx}\to \hat{\bx}+\bd$. Displacements satisfy the relation
\begin{equation}\label{eq:displacement_comm}
    \hat{D}(\bm z)\hat{D}(\bm y)= e^{i\bm z^\top\bm\Omega\bm y}\hat{D}(\bm y)\hat{D}(\bm z).
\end{equation}

In this paper, we consider random Gaussian noise $\be \sim\mathcal{N}(0,{\bV})$ with zero mean and covariance $\bm V$. In which case the channel PDF is a multivariate Gaussian distribution,
\begin{align}
    g[\bV](\be) \coloneqq \frac{1}{(2\pi)^N \sqrt{\det \bV}} \exp\left(-\frac{1}{2}\be^\top \bV^{-1} \be\right),
\end{align}
and we let $\Phi_{\bm V}$ denote the channel. For independent and identically distributed (iid) additive Gaussian noise (AGN) with standard deviation $\sigma$, the covariance matrix is $\bm V=\sigma^2\bm I_{2N}$, and we write the AGN channel simply as $\Phi_{\sigma}$. This is a commonly used, albeit oversimplified, noise model for GKP codes (cf. Refs.~\cite{noh2020o2o, wu2023optimal}). We note that other relevant noise mechanisms, such as photon loss, can be degraded to AGN by a local operation, such as pre-amplification with a quantum limited amplifier~\cite{noh2018GKPcapacity}, though this is not generally optimal for error correction purposes~\cite{zheng2024nearoptimalQEC}.

For a general additive noise channel $\Phi_P$ on $N$ modes, we define the output variance (per mode per quadrature), 
\begin{equation}\label{eq:output_variance}
    \sigma_{\rm out}^2\coloneqq \frac{1}{2N}\int_{\mathbb{R}^{2N}} \diff^{2N}{\bm e} \; P(\bm e) \bm e^\top\bm e.
\end{equation}
For an AGN with covariance $\bm V$, $\sigma^2_{\rm out}=\Tr{\bm V}/(2N)$.

With the relevant Gaussian components introduced, we now move on to non-Gaussian states.
An $N$-mode GKP state $\ket{\Psi}$ represents a $2N$ dimensional (symplectically integral) lattice $\mathcal{L}$ in phase space~\cite{gkp2001}. The lattice can be generated by a set of basis vectors $\{\bm m_j\}$, such that $\bm m_j^\top\bm\Omega\bm m_k\in\mathbb{Z}$ (symplectic integral condition). We package the generators into a generator matrix $\bm M=(\bm m_1, \dots,\bm m_{2N})$. The generators determine the stabilizers of the code via $\hat{S}_j=\hat{D}(\ell \bm m_j)$, where we have set $\ell=\sqrt{2\pi}$ for the fundamental lattice spacing of GKP states. A GKP code state $\ket{\Psi}$ is a +1 eigenstate of all the stabilizers.

A $[[N,k]]$ GKP code, which encodes $k$ qudits into $N$ modes, can be operationally constructed by coupling a set of $k$ independent GKP qudits to $N-k$ canonical ($d=1$) GKP states through a zero-mean Gaussian unitary $\hat{U}_{\rm enc}=\hat{U}_{\bm S,\bm 0}$, where $\bm S$ is a symplectic matrix. The initial generator matrix is $\bm M_{\rm in}= \left(\bigoplus_{j=1}^k\sqrt{d_j}\bm I_2\right)\oplus \bm I_{2(N-k)}$. After encoding, the multi-mode code can be represented by the new generator,
\begin{equation}\label{eq:M_ktoN}
    \bm M_{k\rightarrow N}=\bm S\bm M_{\rm in}.
\end{equation}
which is in one-to-one correspondence with the quantum state in Eq.~\eqref{eq:ktoN_GKP_qudits}. See Appendix~\ref{app:theory_gkp} for a more detailed introduction to the theory of multi-mode GKP codes.

Pauli $X$ and $Z$ operations acting on the initial qudits correspond to displacements along the first $2k$ columns vectors ($\bar{\bm m}_{{\rm in},1}, \bar{\bm m}_{{\rm in},2}, \dots, \bar{\bm m}_{{\rm in},2k}$) of the dual matrix, $\overline{\bm M}_{\rm in}=\bm M_{\rm in}^{-1}$, such that $\hat{X}_{{\rm in},1}=\hat{D}(\ell\bm\bar{\bm m}_{{\rm in},1})$, $\hat{Z}_{{\rm in},1}=\hat{D}(\ell\bm\bar{\bm m}_{{\rm in},2})$ etc. This is consistent with Eq.~\eqref{eq:qudit_paulis}. The dual matrix after encoding is $\overline{\bm M}_{k\rightarrow N}=\bm S \bm M_{\rm in}^{-1}$,
such that $\hat{X}_{k\rightarrow N,1}=\hat{D}(\ell\bm S\bar{\bm m}_{{\rm in},1})$, $ \hat{Z}_{k\rightarrow N,1}=\hat{D}(\ell\bm S\bar{\bm m}_{{\rm in},2})$ etc. 

These Pauli operations are valid logical operations on GKP code states. However, they are not guaranteed to be minimal-weight Paulis. In other words, a Pauli displacement $\bar{\bm m}_J$, such that $\hat{J}=\hat{D}(\ell \bar{\bm m}_J)$ describes some logical Pauli $J$, may not be the smallest possible displacement needed to enact the logical operation. Minimal-weight Paulis are important from an error correction perspective as such determines the smallest displacement required to cause a logical error. This leads us to introduce the GKP Pauli distance~\cite{baptiste2022multiGKP,eisert2022lattice,lin2023closest} for the logical Pauli $J$,
\begin{equation}\label{eq:pauli_dist}
    \mathfrak{D}_J\coloneqq \ell \times \min_{\bm a\in\mathbb{Z}^{2N}}{\norm{\bar{\bm m}_J - \bm M \bm a}}.
\end{equation}
For a square qudit, $\mathfrak{D}_X=\mathfrak{D}_Z=\ell/\sqrt{d}$. The GKP code distance $\mathfrak{D}$ is then defined as the smallest Pauli distance,
\begin{equation}\label{eq:gkp_dist}
    \mathfrak{D}\coloneqq \min_{J}\mathfrak{D}_J.
\end{equation}

GKP states are highly effective at estimating small displacements~\cite{gkp2001,duivenvoorden2017sensor_state}. If a displacement error $\bm e$ happens on an encoded GKP state $\ket{\Psi}$, we measure the stabilizers $\hat{S}_j$ to extract information about $\bm e$ from the error syndrome,
\begin{equation}\label{eq:syndrome}
    \bm s(\bm e) \coloneqq \bm M^{\top}\bm\Omega\bm e\mod\ell,
\end{equation}
where the modulo operation acts element-wise. If desirable, we can perform syndrome-informed counter displacements to correct the error. If the error is small enough, then the error can be corrected with high probability. The syndromes can also be employed to estimate displacements on other modes coupled to the GKP state, such as in oscillators-to-oscillators codes~\cite{noh2020o2o,wu2023optimal}.

We take a simple approach to estimating error displacements that (i) is linear in $\bm s$ and (ii) does not leverage explicit knowledge of the error magnitude (e.g., $\abs{\bm e}\sim\sigma$) in the estimation. More sophisticated strategies can be found in Refs.~\cite{fukui2017PRLanalogQEC,fukui2018PRXftqc,eisert2022lattice,wu2023optimal,lin2023closest,conrad2023ntru}. The generic setting is as follows. Consider two (possibly correlated) random displacements $\bm e_1$ and $\bm e_2$. Suppose $\bm e_2$ occurs on a GKP state, thus permitting stabilizer measurements to extract syndromes $\bm s(\bm e_2)$. The objective is to concoct a good estimator, $\tilde{\bm e}_1$, for $\bm e_1$ based on the syndromes and knowledge about correlations. For simplicity, we employ linear estimation,
\begin{equation}\label{eq:linear_est}
\tilde{\bm e}_1\coloneqq \bm F\bm s(\bm e_2),
\end{equation}
where $\bm F$ is some (possibly rectangular) matrix that is independent of $\bm s$ and $\sigma$.

In the context of recent experimental developments, engineering of single-mode GKP states have been achieved in both microwave and optical domains as well as with trapped ions. In microwave cavity-QED systems, GKP states are generated via coupling a transmon qubit to the cavity modes in a carefully controlled manner~\cite{eickbusch2022fast}. In the optical frequency domain, an itinerant GKP state can be created in a probabilistic, though heralded, fashion with Gaussian boson sampling devices~\cite{sabapathy2019QMLNonGauss,su2019pnrNonGauss,quesada2019NGaussPrep,tzitrin2020ProgressGKP, takase2024GKPgeneration}; see Ref.~\cite{konno2023propagatingGKP} for a proof-of-concept demonstration, albeit far from the quality level to enable break-even error correction. Trapped ion implementations resemble control schemes in cavity-QED mentioned above. However, here, an internal spin state of the ion is used to control its motional degrees of freedom~\cite{home2019GKPnature,home2022QECgkp}. To develop multi-mode GKP codes, one can envisage starting with standardized ``factories'' that produce single-mode square-GKP qudits through local bosonic control schemes or heralded methods. Equipped with universal Gaussian operations, one can then, in principle, entangle these factory qudits to create an arbitrary multi-mode GKP code [per Eq.~\eqref{eq:M_ktoN}].


\section{Distributed Two-Mode Squeezing Codes}\label{sec:dtms_codes}

In this section, we formally introduce the dtms-GKP codes, including examples of single-qudit, oscillators-to-oscillators, and two-qubit codes. Using the generator matrix of the dtms-GKP qubit codes, we also present numerical results for the code distance.

\subsection{Setups and code definitions}\label{ssec:dtms_setup}

We have presented our unitary Gaussian encoder for the dtms code, $\hat{U}_{\rm dtms}$ in Eq.~\eqref{eq:unitary_dtms}, which applies to both qudit-to-oscillators codes and oscillators-to-oscillators codes. For the majority of this paper, we focus on encoding a single data mode ($k=1$) into $N$ modes, where $N-1$ modes consist of canonical GKP anillae. The dtms encoder $\hat{U}_{\rm dtms}$ [see also Eq.~\eqref{eq:unitary_dtms} \QZ{and Fig.~\ref{fig:dtms_schematic}(c)}] can be represented by a symplectic matrix,
\begin{equation}\label{eq:dtms_encoder}
        \bm S_{\rm dtms}=\left(\bm S_{G}\oplus\bm I_{2(N-2)}\right)\left(\bm I_{2}\oplus\bm B\right).
    \end{equation}
The encoder consists of (1) a two-mode squeezer, $\bS_{G}$, with tunable gain $G$ that couples the lone data mode to a single ancilla, and (2) a configurable multi-port interferometer, $\bB$, that mixes the ancillary GKP modes. For iid noise, we find it sufficient to further reduce the multi-port interferometer $\bm B$ to a cascaded staircase of beamsplitters with transmission probabilities (from top to bottom) $\eta_{1}=1/(N-1)$, $\eta_{2}=1/(N-2)$, \dots, $\eta_{N-2}=1/2$, and a single-phase, $\phi$, [see Fig.~\ref{fig:dtms_schematic}(c) and Appendix~\ref{app:beamsplitter}]
\begin{equation}\label{eq:balanced_B}
\renewcommand{\arraystretch}{1}
    \bm B=
    \left(
    \begin{array}{c c | c c}
        \frac{\bm{R}(\phi)}{\sqrt{N-1}} & & & \\ 
        \vdots & & & \bm B^\prime \\
        \frac{\bm{R}(\phi)}{\sqrt{N-1}} & & & 
    \end{array}
    \right)
\end{equation}
where $\bm R(\phi)$ is a $2\times2$ rotation matrix [Eq.~\eqref{eq:phase}], and $\bm B^\prime$ is a $2N\times 2(N-1)$ sub-block that does not appreciably impact the code. The above construction ensures that the initial correlations (generated by two-mode squeezing) between the lone data and single ancilla is distributed uniformly to all the ancillary modes. While the single phase, $\phi$, allows us just enough freedom to balance the code (see below for details). We now formally define our codes, starting with the dtms qudit code.

\begin{definition}[dtms-GKP Qudit Encoding]\label{def:dtms_qudit}
    Consider a square GKP qudit of dimension $d$ and $N-1$ canonical GKP ancillae. The initial generator matrix is given by a trivial direct sum, $\bm M_{\rm in}=\left(\sqrt{d}\bm I_{2}\oplus\bm I_{2(N-1)}\right)$. We define a dtms-GKP qudit code by the resulting generator matrix,
    \begin{equation}
        \bm M_{{\rm dtms}}\coloneqq \bm S_{\rm dtms}\bm M_{\rm in},
    \end{equation}
    where $\bm S_{\rm dtms}$ is written explicitly in Eq.~\eqref{eq:dtms_encoder}. Logical Pauli operations can be associated with the first $2$ columns of the dual matrix, 
    \begin{equation}\label{eq:dual_dtms}
        \overline{\bm M}_{\rm dtms}=\bm S_{\rm dtms}\left(\bm I_{2}/\sqrt{d}\oplus\bm I_{2(N-1)}\right).
    \end{equation}
\end{definition}
\QZ{ The columns of $\bm{M}_{\rm dtms}$ describe the symmetry translations of the $2N$-dimensional lattice in real phase space, $\mathcal{L}$, that is associated with the dtms-GKP qudit code space, $\mathcal{C}$. By definition, a logical dtms-GKP qudit state $\ket{\Psi_L}\in\mathcal{C}$ is a +1 eigenstate of unitary displacements (stabilizers) along the column vectors of $\bm M_{\rm dtms}$. For further examples, please refer to Definition~\ref{def:dtms_2qubit} and Definition~\ref{def:dtms_o2o}. For a concrete example, a two-mode squeezed GKP qudit state can be written in similar form as Eq.~\eqref{eq:ktoN_GKP_qudits}, namely,
\begin{equation}
    \ket{\Psi_L}=\hat{U}_{\bm S_G}\left(\ket{\Psi}\otimes\ket{\rm GKP}\right),
    \label{eq:logical}
\end{equation}
where $\hat{U}_{\bm S_G}$ is a unitary two-mode squeezing operation (see Eq.~\eqref{eq:S_TMS}), $\ket{\Psi}=\sum_j\alpha_j\ket{j;d}$ is an arbitrary single-mode square GKP qudit with basis states $\{\ket{j;d}\}_{j=1}^d$ defined in Eq.~\eqref{eq:gkp_comp_basis}, $\ket{\rm GKP}=\ket{0;1}$ is a canonical GKP state with trivial dimension ($d=1$), and $\ket{\Psi_L}$ is the two-mode logical representation of $\ket{\Psi}$, respectively. The generator matrix is simply $\bm M_{\rm dtms}=\bm S_G\left(\sqrt{d} \bm I_2 \oplus \bm I_2\right)$ for this code space.
}

Due to the structure of the code, it is easy to infer an upper bound on the code distance directly from the dual matrix~\eqref{eq:dual_dtms}. The first two columns of $\overline{\bm M}_{\rm dtms}$ are Pauli vectors for logical $X$ and $Z$ operations of length $\sqrt{2G-1}\mathfrak{D}_{\square}(d)$, where $\mathfrak{D}_{\square}(d)=\sqrt{2\pi/d}$ is the local distance for the square qudit, and we have incorporated the lattice spacing $\ell=\sqrt{2\pi}$. However, these are not necessarily minimal-weight Paulis, \QZ{ with minimal-weight Paulis defined as the smallest displacements that induce logical operations [see Eq.~\eqref{eq:pauli_dist}].} Therefore, \QZ{ by Eq.~\eqref{eq:gkp_dist}},
\begin{equation}\label{eq:dtms_distance}
    \mathfrak{D}_{\rm dtms}\leq \sqrt{2G-1}\mathfrak{D}_\square(d).
\end{equation}
Since the gain is a continuously tunable parameter, we imagine starting from $G=1$ (no coupling) and slowly increasing the gain ($G>1$) to increase the code distance, such that the above inequality is always met. Alas, this continuous improvement cannot be sustained indefinitely. There will be some optimal value, $G^\star$, that saturates the inequality above and maximizes the code distance, \QZ{ above which the distance will actually decrease. See Fig.~\ref{fig:distance_contour} in Section~\ref{ssec:numerics} for numerical evidence of this intuition.}

Regarding the single phase, $\phi$, in the interferometer $\bm B$ [Eq.~\eqref{eq:balanced_B}], for $\phi=0$, the dtms qudit code is of Calderbank-Shor-Steane (CSS) type and $\mathfrak{D}_X=\mathfrak{D}_Z=\mathfrak{D}_Y/\sqrt{2}$. However, \QZ{ we have found that} keeping one phase as a free parameter allows us just enough freedom to balance the code, such that $\mathfrak{D}_X\approx\mathfrak{D}_Y\approx\mathfrak{D}_Z$ for some optimal set $(G^\star,\phi^\star)$, and, consequently, increase the code distance beyond what is achievable with the CSS-type counterpart ($\phi=0)$. \QZ{ Intuitively, the balanced condition ($\mathfrak{D}_X\approx\mathfrak{D}_Y\approx\mathfrak{D}_Z$) is met by gradually increasing the $X$ and $Z$ distances while decreasing the $Y$ distance, until equality is met.} We find $G^\star$ and $\phi^\star$ numerically (see Section~\ref{ssec:numerics} for details).

The current formulation of the dtms qudit code (Definition~\ref{def:dtms_qudit}) is designed to safeguard a single qudit. To extend this construction to $k$ qudits, a straightforward approach is to introduce $k$ two-mode squeezers, assigning one to each qudit. This limits the number of active elements to scale with the number of qudits ($\sim k$) as opposed to directly scaling with the size of the code block ($\sim N$). See Section~\ref{sec:discussion} and Fig.~\ref{fig:dtms_kmodes} for further discussion. However, there may exist more clever ways to increase the encoding rate ($k/N$) and code distance without introducing more active elements. We demonstrate this with a two-qubit code example.

\begin{figure}
    \centering
    \includegraphics[width=.6\linewidth]{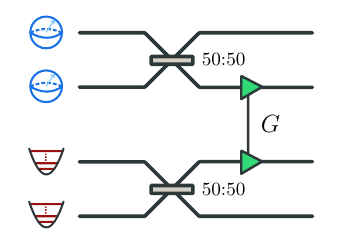}
    \caption{Schematic of the encoder for a dtms-GKP two-qubit code $(N=4, k=2)$.}
    \label{fig:dtms42_encoder}
\end{figure}

\begin{definition}[dtms Two-Qubit Encoding]\label{def:dtms_2qubit}
    Consider two square GKP qubits and $N-2$ canonical GKP ancillae. The initial generator matrix is given by a trivial direct sum, $\bm M_{\rm in}=\left(\sqrt{2}\bm I_{4}\oplus\bm I_{2(N-2)}\right)$. The encoder for the two-qubit dtms-GKP code is specified by the symplectic matrix, 
    \begin{equation}\label{eq:2qubit_encoder}
        \bm S_{{\rm dtms},2}= \left(\bm I_2\oplus \bm S_G\oplus \bm I_{2(N-3)}\right)\left(\bm B_{50:50}\oplus \bm B\right),
    \end{equation}
    where $\bm B_{50:50}$ represents a 50:50 beamsplitter and $\bm B$ is the multi-port interferometer of Eq.~\eqref{eq:balanced_B}. The generator matrix for the code is $\bm M_{{\rm dtms}; 2}=\bm S_{{\rm dtms},2}\left(\sqrt{2}\bm I_4\oplus\bm I_{2(N-2)}\right)$.
\end{definition}
The dtms two-qubit codes rely on (1) a single two-mode squeezer of gain $G$ that couples one of the qubits to one of the ancillary GKP, (2) a 50:50 beamsplitter that entangles the data qubits, and (3) a multi-port interferometer that entangles the ancillary GKP modes. See Fig.~\ref{fig:dtms42_encoder} for an illustration of the encoder for $N=4$. Analogous to the dtms single-qudit code, we find an upper bound on the code distance,
\begin{equation}\label{eq:2qubit_distance}
    \mathfrak{D}_{{\rm dtms},2}\leq \sqrt{G}\mathfrak{D}_{\square}.
\end{equation}

Finally, we encode an oscillator into many oscillators. The dtms-O2O codes (defined formally below) are, in many respects, similar to the dtms qudit codes because both codes utilize the same (unitary Gaussian) encoder $\hat{U}_{\rm dtms}$ [see Eq.~\eqref{eq:unitary_dtms} and Eq.~\eqref{eq:dtms_encoder}]. One major difference,\footnote{Another difference is the actual value of the gain $G$ preferred by each code.} however, is that the data mode for the dtms-O2O code is an analog quantum state (e.g., squeezed vacuum) that does not have a lattice-type description. Thus, rather than a generator matrix, of importance to the O2O code is the correlated noise matrix, $\bm V$~\cite{noh2020o2o,wu2023optimal}, which determines the analog output noise of the code. The noise matrix comes from sandwiching the (iid) AGN channels, $\Phi_{\sigma}$, between the encoder, $\mathcal{U}_{\rm enc}$, and (the unitary part of) the decoder, $\mathcal{U}_{\rm dec}=\mathcal{U}_{\rm enc}^{-1}$. From this we define our code.

\begin{figure*}
    \centering
    \includegraphics[width=.85\linewidth]{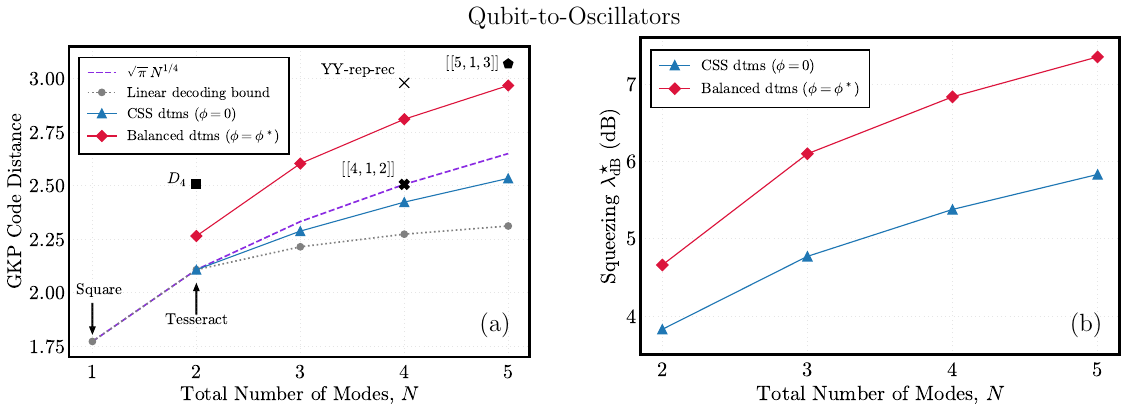}
    \caption{(a) Code distance $\mathfrak{D}_{\rm dtms}$ for dtms qubit code. Assorted codes from the literature are shown for reference: Square, Tesseract~\cite{baptiste2022multiGKP}, \QZ{ $D_4$},  Square-$[[4,1,2]]$, \QZ{ YY-rep-rec~\cite{lin2023closest}}, Square-$[[5,1,3]]$, and scaling of GKP-surface--type codes $\sim N^{1/4}$. Linear decoding bound inferred from two-stage linear decoder [Section~\ref{sec:decoding_dtms}, Eq.~\eqref{eq:D_N}]. (b) Optimized squeezing for dtms qubit code. Two-mode squeezing gain $G^\star$ converted to equivalent single-mode squeezing $\lambda_{\rm dB}^\star\coloneqq 20\log_{10}(\sqrt{G^\star}+\sqrt{G^*-1})$.}
    \label{fig:code_distance}
\end{figure*}

\begin{definition}[dtms-O2O Encoding]\label{def:dtms_o2o}
The dtms-O2O code encodes a single-mode continuous-variable (CV) state into $N$ modes, where $N-1$ modes are canonical GKP ancillae. The unitary encoder, $\hat{U}_{\rm dtms}$, is specified by the symplectic matrix $\bm S_{\rm dtms}$ [Eq.~\eqref{eq:dtms_encoder}]. Assuming iid AGN of variance $\sigma^2$, the noise channel $\Phi_{\sigma}$ transforms under the unitary encoding-decoding pair as,
    \begin{equation}
    \Phi_{\bm V}=\mathcal{U}_{\rm dtms}^{-1}\circ\Phi_{\sigma}\circ\mathcal{U}_{\rm dtms},
\end{equation}
where $\mathcal{U}_{\rm dtms}(\cdot)=\hat{U}_{\rm dtms}(\cdot)\hat{U}_{\rm dtms}^\dagger$ and $\Phi_{\bm V}$ is a correlated AGN channel with covariance $\bm V$. Explicitly,
\begin{equation}\label{eq:dtms_corrV}
    \bm V=\sigma^2\bm S_{\rm dtms}^{-1}\bm S_{\rm dtms}^{-\top}.
\end{equation}
\end{definition}
The correlations between the data mode and the ancillary GKP modes (generated by the unitary interaction $\hat{U}_{\rm dtms}$) allow for an estimate of the data noise via stabilizer measurements on the ancillary GKP modes. In turn, these estimates can be used in the measurement-stage of decoding (see Section~\ref{sec:decoding_dtms}) to perform counter displacements on the data mode, thereby suppressing the analog noise on the data.

\QZ{
\subsubsection{Connection to distributed quantum sensing}
\label{sec:connection}
}

\QZ{As shown in Fig.~\ref{fig:dtms_schematic}(b), a continuous-variable distributed quantum sensing protocol~\cite{zhuang2018dqs} prepares the CV multi-partite entangled probe state by distributing a single-mode squeezed state to multiple sensor nodes through a beamsplitter network. The symplectic matrix, $\bm S_{\rm DQS}$ (associated with a Gaussian unitary, $\hat{U}_{\bm{S}_{\rm DQS}}$), that describes the state preparation is given by 
\begin{equation}
{\bm S}_{\rm DQS}= \bm B \bm S^{(1)}_G,
\end{equation} 
where $\bm S^{(1)}_G$ is defined in Eq.~\eqref{eq:S_SMS}. Such an encoding allows the global reduction of vacuum fluctuations via multi-mode squeezing. As shown in Fig.~\ref{fig:dtms_schematic}(c), the dtms encoding operation $\bm S_{\rm dtms}$ first applies the beamsplitter transform and then performs two-mode squeezing, as given in Eq.~\eqref{eq:dtms_encoder}. The decoding process starts with the reverse unitary, $\bm S_{\rm dtms}^{-1}=\left(\bm I_{2}\oplus\bm B^T\right)\left(\bm S_{-G}\oplus\bm I_{2(N-2)}\right)$. Compared with the state preparation operation ${\bm S}_{\rm DQS}$, the dtms operation $\bm S_{\rm dtms}^{-1}$ inherits the same spirit of distributing squeezing, with the exception of replacing the single-mode squeezer for distributed quantum sensing with a two-mode squeezer for error correction. In the dtms code, the two-mode squeezing correlates the data mode error and the ancilla error; then the beamsplitters distribute the correlations to multiple GKP ancillae, such that the error on each GKP ancilla is small (compared to GKP grid size) and can be estimated with high precision. 
}

\subsection{Numerical results for code distance}\label{ssec:numerics}

We present our results for the GKP code distance of dtms codes obtained through numerical optimization. On the one hand, the code distance serves as a valuable first-order metric for evaluating the performance of GKP codes. On the other hand, it is worth noting that this optimization is agnostic to the details of the error channel that the code is meant to correct for or the decoder used in the error correction process. 

The basic idea behind our numerical optimization routine is as follows: Given a code (e.g., the dtms qudit code of Definition~\ref{def:dtms_qudit}) specified by parameters $\bm\vartheta$ (e.g., gain $G$, beamsplitter transmissivities $\eta_j$, and phases $\phi_j$), we numerically optimize the parameter set to maximize the GKP code distance, i.e. $\bm\vartheta^\star={\rm argmax}_{\bm\vartheta}~\mathfrak{D}(\bm\vartheta)$,
thus determining the optimal set $\bm\vartheta^\star$ and the corresponding maximum code distance $\mathfrak{D}(\bm\vartheta^\star)$.

The code distance is numerically calculated by searching for the closest lattice point that minimizes the GKP code distance, as defined in Eqs.~\eqref{eq:pauli_dist} and~\eqref{eq:gkp_dist}. 
\QZ{
From Eq.~\eqref{eq:dtms_encoder}, we first determine the generator matrix $\bM$ and its dual matrix $\overline{\bm M}$, which represent logical Pauli operators. Then the Lenstra-Lenstra-Lov\'{a}sz (LLL) lattice reduction algorithm \cite{Lenstra1982-ks} is applied to $\bm M$ to obtain a reduced basis of the generator matrix. This algorithm runs in polynomial time, and the reduced basis $\{\boldsymbol{m}_i\}_{i=1}^{2N}$, which are column vectors of $\bm M$, is more compact than the original basis. Next we take a vector $\bm{v}$ in the dual matrix $\overline{\bm M}$ that corresponds to a logical Pauli in the set $\{X,Y,Z\}$ to calculate its distance with respect to the dual lattice.
Babai's algorithm \cite{Babai1986-cy} is used here to find a lattice point $\boldsymbol{m}_L$ in the dual lattice that is closest to the given vector $\boldsymbol{v}$. However, Babai's algorithm might not return the true closest point, as the closest point problem is NP-hard. Therefore, the lattice point returned by Babai's algorithm simply serves as a strong candidate. Then a brute-force search with cutoff $l$ is conducted around the candidate point $\boldsymbol{m}_L$ as
\begin{align}
    \min_{-l\leq k_i \leq l} \norm{\sum_{i=1}^{2N} \boldsymbol{v}-\boldsymbol{m}_L+k_i \boldsymbol{m}_i}. \label{eq:brute-force-algorithm}
\end{align}
After the brute-force search, we obtain the closest distance from the logical Pauli to all lattice points around the candidate point. A total of $(2N)^{2l}$ points are considered, and we take $l\leq 4$ since the number of points diverges quickly with $N$.
We perform the above numerical search for both dtms qubit codes (Definition~\ref{def:dtms_qudit} with $d=2$) and dtms two-qubit codes (Definition~\ref{def:dtms_2qubit}).
We validate our optimized results by calculating the distance in Eq.~\eqref{eq:brute-force-algorithm} with $l$ increased by one, ensuring that the distance does not change with more points.
}

\subsubsection{Example 1: Single-qubit codes}

We first consider a CSS-type code with no phases ($\phi_j=0$). In this scenario, the $2N$-dimensional lattice decomposes into two $N$-dimensional sublattices. Consequently, the generator matrix can be divided into independent $q$ and $p$ sections, such that $\mathfrak{D}_X = \mathfrak{D}_Z = \mathfrak{D}_Y / \sqrt{2}$ and thus $\mathfrak{D}=\mathfrak{D}_X$. The distance $\mathfrak{D}_X$ is computed numerically for the $q$ section. By optimizing the beamsplitter array, $\eta_i$, and the two-mode squeezing gain, $G$, with a cutoff $l = 4$, we generally need to maximize the code distance over $N-1$ parameters. This optimization is further verified by ensuring that the code distance remains consistent when the cutoff is increased to $l = 5$. \QZ{Note that since the problem is NP-hard, the distance we provide will be approximate solutions, similar to Ref.~\cite{lin2023closest} and others.}

However, we verify the intuition that, for both CSS- and non-CSS--type codes, the optimal values for the beamsplitter parameters are $\eta_1 = 1/\sqrt{N-1}$, ..., $\eta_{N-2} = 1/\sqrt{2}$. This is confirmed in numerical (local) optimization with three and four modes, where random initialization of beamsplitter parameters converge to the above optimal values. This suggests that a uniform beamsplitter array outperforms other beamsplitter configurations. This is expected on symmetry grounds, as no single mode should be preferred over another. Therefore, correlations generated from the two-mode squeezing interaction in the dtms code should be uniformly shared by all ancillary modes, a result guaranteed by the aforementioned parameter choices. Thus, for our CSS-type codes, only one parameter, $G$, needs to be optimized. Numerical results of the code distance for CSS-type dtms qubit codes are presented in Fig.~\ref{fig:code_distance}(a) as blue triangles.

\begin{figure*}
    \centering
    \includegraphics[width=.85\linewidth]{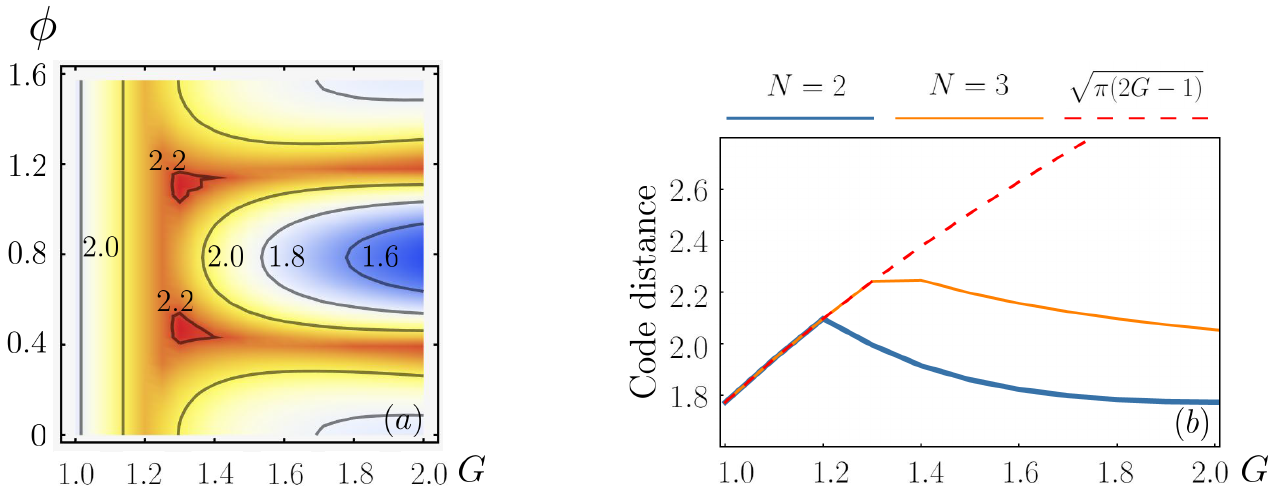}
    \caption{\QZ{ (a) Contour of code distance $\mathfrak{D}_{\rm dtms}$ for $N=2$ versus phase $\phi$ and gain $G$. (b) Line plot of code distance with respect to gain for $N=2,3$ modes.} }
    \label{fig:distance_contour}
\end{figure*}

Next, we incorporate $N-1$ phase rotations, $\phi_j$, into the uniform beamsplitter array in order to balance the code, such that $\mathfrak{D}_X=\mathfrak{D}_Y=\mathfrak{D}_Z$. The resulting code (with non-trivial phases) is a non-CSS--type code that correlates the $q$ and $p$ sections, leading to a genuine $2N$-dimensional lattice. Importantly, balancing the code further enhances the code distance beyond the CSS-type counterpart. The intuition is as follows. Recall that we compute the code distance via $\min \{\mathfrak{D}_X, \mathfrak{D}_Y, \mathfrak{D}_Z\}$. Compared to the CSS-type code ($\phi=0$), phase rotations allow us to increase $\mathfrak{D}_X$ and $\mathfrak{D}_Z$ while decreasing $\mathfrak{D}_Y$. Therefore, we imagine starting with zero phases and slowly tuning them until the Pauli distances converge, effectively increasing the overall code distance. 

Our numerical calculations (with two to five modes) suggest that it is sufficient to choose a single phase, $\phi\coloneqq \phi_1 = ... = \phi_{N-1}$, to balance the code. Hence, in our final numerical search for optimal codes, only two parameters, $G$ and $\phi$, are optimized. As a technical note, the code distance remains unchanged when $\phi_i \rightarrow \phi_i + \pi/2$. Therefore, we restrict the range of $\phi_i$ to $0 \leq \phi_i < \pi/2$ in numerical calculations. The results of our numerical optimization for non-CSS--type dtms qubit codes are depicted in Fig.~\ref{fig:code_distance}(a) as red diamonds. We further verify that the code distance, for both CSS- and non-CSS-type codes, obeys the relation $\mathfrak{D} =\sqrt{(2G^\star-1)\pi}$, as expected~\eqref{eq:dtms_distance}.

\QZ{ For reference, we plot contours of the GKP code distance in parameter space $(G,\phi)$ for a two-mode code in Fig.~\ref{fig:distance_contour}(a), illustrating the change in the code distance with the gain $G$ and also the periodicity in $\phi$. Scaling of the code distance with gain for $N=2$ and $N=3$ mode codes is shown in Fig.~\ref{fig:distance_contour}(b). Results agree with the predicted code distance in Eq.~\eqref{eq:dtms_distance} below the optimal gain value, $G^\star$. For $G>G^\star$,  the actual code distance decreases, indicating that the distance cannot increase indefinitely with the gain for fixed number of modes, $N$.}

\subsubsection{Example 2: Two-qubit codes}

We focus on CSS-type two-qubit codes of $N=3$ and $N=4$ modes, as generically defined in Definition~\ref{def:dtms_2qubit}. The $N=3$ two-qubit code is obtained by entangling two GKP qubits by a 50:50 beamsplitter and then coupling one qubit to a single canonical GKP state via two-mode squeezing. Logical operations are specified by the dual matrix,
\begin{equation}\label{eq:dual_2dtms3}
        \overline{\bm M}_{{\rm dtms}, 2}^{(3)}=
        \begin{pmatrix}
        \renewcommand*{\arraystretch}{1}
            \frac{1}{2}\bm I_2 & - \frac{1}{2}\bm I_2 & \bm 0 \\[.25em]
             \frac{\sqrt{G}}{2}\bm I_2 & \frac{\sqrt{G}}{2}\bm I_2 & \sqrt{G-1}\bm Z \\[.5em] 
             \frac{\sqrt{G-1}}{2} \bm Z &  \frac{\sqrt{G-1}}{2} \bm Z & \sqrt{G} \bm I_2  
        \end{pmatrix}.
    \end{equation}

The $N=4$ two-qubit code is generated by independently entangling two GKP qubits and two canonical GKP ancilla via 50:50 beamsplitters. Subsequently, one GKP qubit and one GKP ancilla are then coupled via two-mode squeezing. See Fig.~\ref{fig:dtms42_encoder} for an illustration. 
Logical operations are specified by the dual matrix, 
    \begin{equation}\label{eq:dual_2dtms4}
        \overline{\bm M}_{{\rm dtms}, 2}^{(4)}=
        \begin{pmatrix}
        \renewcommand*{\arraystretch}{1}
            \frac{1}{2}\bm I_2 & - \frac{1}{2}\bm I_2 & \bm 0 & \bm 0 \\[.25em]
             \frac{\sqrt{G}}{2}\bm I_2 & \frac{\sqrt{G}}{2}\bm I_2 & \sqrt{\frac{G-1}{2}}\bm Z & -\sqrt{\frac{G-1}{2}}\bm Z \\[.5em] 
             \frac{\sqrt{G-1}}{2} \bm Z &  \frac{\sqrt{G-1}}{2} \bm Z & \sqrt{\frac{G}{2}} \bm I_2 & -\sqrt{\frac{G}{2}} \bm I_2 \\[.5em]
            \bm 0 & \bm 0 & \frac{1}{\sqrt{2}}\bm I_2 & \frac{1}{\sqrt{2}}\bm I_2
        \end{pmatrix}.
    \end{equation}
For optimal performance, we numerically optimize the gain, $G$, of the two-mode squeezer, following the same approach as in the single-qubit examples.

We find that the code distance is determined by the optimal gain $G^\star$ via $\mathfrak{D}_{{\rm dtms},2} = \sqrt{G^\star\pi}$, as expected from Eq.~\eqref{eq:2qubit_distance}. Numerical analysis reveals that $G^\star=4/3$ and $G^\star=2$ for $N=3$ and $N=4$, respectively, implying that $\mathfrak{D}_{{\rm dtms},2}^{(3)}=\sqrt{4\pi/3}$ and $\mathfrak{D}_{{\rm dtms},2}^{(4)}=\sqrt{2\pi}$. The $N=3$ code is intriguing, enabling the encoding of two qubits into three modes while achieving a code distance that is $2/\sqrt{3}\approx 1.15$ times larger than the square code. Curiously, the distance for the $N=4$ code coincides with the standard $[[4,2,2]]$ code. We emphasize that a simple concatenation of square GKP qubit and the qubit $[[4,2,2]]$ code will need five CNOT gates~\cite{roffe2019quantum}, leading to five active Gaussian operations (SUM-gate), while our GKP-dtms code achieves the same code distance with only a single active Gaussian operation.

\subsubsection{Comparison to prior works}
To end this section, we briefly contextualize our results for code distance with recent literature. Regarding the code distance for single-qubit dtms codes, Lin et al~\cite{lin2023closest} performed an unstructured numerical optimization over all symplectic matrices $\bS \in \Sp(2N)$, which have (in the symmetry-reduced setting) $N^2+N$ free parameters. The authors successfully identified effective codes. However, due to their unstructured nature, these codes practically need to be constructed through a Bloch-Messiah decomposition, which requires $N$ active squeezers and a general $N$-port interferometer. Contrariwise, our dtms-GKP code designs, based on a structured beamsplitter array and having only two parameters, circumvent such complexities while achieving almost equivalent performance as the optimized codes found in Ref.~\cite{lin2023closest}. \QZ{ On the other hand, the authors of Ref.~\cite{lin2023closest} also argued the existence of the so-called YY-rep-rec code, with distance $\mathfrak{D}_{\rm YY}=\sqrt{2\pi}(N/2)^{1/4}$, which demonstrates a constant-factor improvement over the square GKP surface code (see Fig.~\ref{fig:code_distance}). Curiously though, the numerically optimized codes considered in that work do not achieve the distance of the theoretically proposed YY-rep-rec code.} Regarding other prototypical codes, although our results do not surpass the perfect [[5,1,3]] code, our approach comes remarkably close, differing by $\lesssim .1$ in the code distance (see Fig.~\ref{fig:code_distance}), while only requiring a single active component. Additionally, our balanced code designs exhibit larger distances compared to the recently discovered two-mode Tesseract qubit code~\cite{baptiste2022multiGKP} and the standard [[4,1,2]] code. 


\section{Two-Stage Decoder for dtms Codes} \label{sec:decoding_dtms}

\begin{figure}
    \centering
    \includegraphics[width=\linewidth]{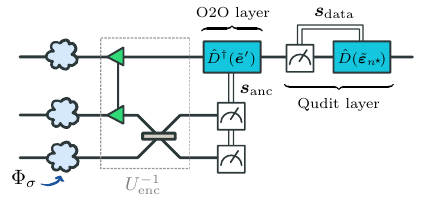}
    \caption{Schematic of the two-stage decoder for dtms-GKP codes.  $\bm s_{\rm anc}$ and $\bm s_{\rm data}$ denote syndrome information from stabilizer measurements at the O2O layer and qudit layer, respectively. $\tilde{\bm e}^\prime$ denotes a syndrome-informed estimator at the O2O layer. $\tilde{\bm\varepsilon}_{\bm n^\star}$ denotes syndrome-informed estimator at the qudit layer, used to displace the local GKP state back to the nearest integer $n^\star$. For analog data, only the O2O layer is necessary. A $N=3$ dtms code is shown for concreteness.}
    \label{fig:linear_decoder}
\end{figure}

Optimal decoding of lattice codes is generally a computationally hard problem~\cite{conway2013sphere,eisert2022lattice,lin2023closest,conrad2023ntru}. To bypass this complexity and shed light on the potential performance of dtms codes, we opt for a simpler approach: We propose a two-stage linear decoder that first decodes at the oscillators-to-oscillators layer then at the local qudit layer. See Fig.~\ref{fig:linear_decoder} for an illustration. Though the decoder is, in general, not optimal, it allows straightforward analytical results and performs quite well for small code sizes, with results indicating near-optimal performance for the $N=2$ dtms qubit code. Investigation into improved decoders---better designed to leverage the specific structure of dtms codes---is left to future study.

Here we summarize our two-stage decoder, providing some mathematical details but deferring derivations to Appendix~\ref{app_ssec:asymptotics}. The unitary component of the decoder, $U_{\rm enc}^{-1}$, correlates the data and ancillae displacement noises (clouds in Fig.~\ref{fig:linear_decoder}) via two-mode squeezing. Beamsplitters then distribute the locally amplified noise uniformly to the ancillary modes. The first measurement step of decoding begins at the oscillators-to-oscillators level (O2O layer). Stabilizer measurements are conducted on the GKP ancillae and syndrome-informed counter displacements are applied on the data to locally reduce the data noise. If the data hold discrete information (i.e., a qudit), then local qudit decoding (Qudit layer) is subsequently performed. Conditioned on the residual displacements from the first stage, the qudit stage involves local GKP stabilizer measurements to acquire an estimate of the residual noise. This estimate is used to then shift back to the nearest lattice point of the local qudit code, analogous to standard single-qubit GKP decoders~\cite{gkp2001,noh2020surface}.

Let $\bm e\sim\mathcal{N}(0,\sigma^2\bm I_{2N})$ be the initial displacement noises on the modes and $K=N-k$ be the number of ancillary GKP modes. Following the unitary component of the decoding map, the noises are correlated and described by $\bm e^\prime\oplus\bm e_{\rm anc}^\prime=\bm S_{\rm enc}^{-1} \bm e\sim\mathcal{N}(0,\bm V)$, where $\bm V=\sigma^2\bm S_{\rm enc}^{-1}\bm S_{\rm enc}^{-\top}$, $\bm S_{\rm enc}$ represents the encoder, and we have written the displacement as a direct sum of data noises, $\bm e^\prime$, and ancillary noises, $\bm e_{\rm anc}^\prime$. For dtms qudit codes and O2O codes, $\bm S_{\rm enc}=\bm S_{\rm dtms}$ of Eq.~\eqref{eq:dtms_encoder}. Whereas, for dtms two-qubit codes, $\bm S_{\rm enc}=\bm S_{{\rm dtms},2}$ of Eq.~\eqref{eq:2qubit_encoder}. Performing stabilizer measurements on the ancillary GKP modes in the O2O decoding stage, we extract error syndromes $\bm s_{\rm anc}=\bm\Omega\bm e_{\rm anc}^\prime \mod\ell$. From the ancillary syndromes, we estimate the noises on the data via linear estimation,
\begin{equation}
    \Tilde{\bm e}^\prime=\bm F\bm\Omega^\top\bm s_{\rm anc},
\end{equation}
where $\bm F$ is a $2k\times 2K$ rectangular matrix that is independent of $\bm s_{\rm anc}$ and $\sigma$. For (CSS-type) dtms qudit codes and O2O codes, we choose
\begin{equation}
    \bm F_{\rm dtms}=\frac{-C_G}{\sqrt{K}}\left(\bm Z\, \bm Z\, \dots\, \bm Z\right),
\end{equation}
where
\begin{equation}\label{eq:c_constant}
    C_G=2\frac{\sqrt{G(G-1)}}{2G-1}.
\end{equation}
For dtms two-qubit codes, we choose
\begin{equation}
    \bm F_{{\rm dtms},2}=\frac{-C_G}{\sqrt{2K}}\begin{pmatrix}
        \bm Z & \bm Z & \dots & \bm Z\\
        \bm Z & \bm Z & \dots & \bm Z
    \end{pmatrix}.
\end{equation}
Intuitively, these choices minimize the variance of the residual noise $\bm\varepsilon=\bm e^\prime-\Tilde{\bm e}^\prime$ in a first-order (Gaussian) approximation. \QZ{ This is analogous to the original O2O code proposed in Ref.~\cite{noh2020o2o} but extended to a multi-mode distributed scenario. For further mathematical explanation of this specific O2O decoder, we refer the reader to Appendix~\ref{app_ssec:asymptotics}.}

After a counter displacement, $\hat{D}(-\Tilde{\bm e}^\prime)$, in the O2O layer of error correction, the resulting output distribution of the residual error, $\bm\varepsilon$, is a non-Gaussian PDF,
\begin{equation}\label{eq:o2o_pdf}
    P(\bm\varepsilon)=\sum_{\bm n\in\mathbb{Z}^{2K}}f(\bm n) \times g[\bm\Sigma]\big(\bm\varepsilon-\bm\mu({\bm n})\big),
\end{equation}
where $\bm\Sigma$ is the conditional data covariance and ${\bm\mu(\bm n)=\ell\bm F\bm\Omega^\top\bm n}$ is a discretized mean that encodes lattice effects from the ancillary GKP. Their explicit forms vary depending on the encoding. Let $n_q\coloneqq\sum_{i=\text{odd}}n_i$ and $n_p\coloneqq\sum_{i=\text{even}}n_i$, where $n_i$ is the $i$-th component of $\bm n\in\mathbb{Z}^{2K}$ in the summation. For single-qudit codes and O2O codes, we can then write covariance and discrete mean as
\begin{align}\label{eq:dtms_sigma}
    \bm\Sigma_{\rm dtms}&= \left(\frac{\sigma^2}{2G-1}\right)\bm I_2 \\
    \qq{and} \bm\mu_{\rm dtms}^\top(\bm n)&=\frac{\ell C_G}{\sqrt{K}}(n_p, n_q). \label{eq:dtms_mu}
\end{align}
Whereas, for two-qubit codes,
\begin{align}\label{eq:2qubit_sigma}
    \bm\Sigma_{{\rm dtms},2}&= \left(\frac{\sigma^2 G}{2G-1}\right)\bm I_2 \\
    \qq{and} \bm\mu_{{\rm dtms},2}^\top(\bm n)&=\frac{\ell C_G}{\sqrt{2K}}
    (n_p, n_q, n_p, n_q).\label{eq:2qubit_mun}
\end{align}
where the constant $C_G$ is defined in Eq.~\eqref{eq:c_constant}.

The quantity $f(\bm n)$ in Eq.~\eqref{eq:o2o_pdf} is a discrete probability, such that $0\leq f(\bm n)\leq 1$ and $\sum_{\bm n\in\mathbb{Z}^{2K}}f(\bm n)=1$. Formally,
\begin{equation}\label{eq:f_n}
    f({\bm n})\coloneqq\int_{\bm x\in\mathscr{I}_{\bm n}(\ell)}\dd{\bm x}g[\bm\Sigma_{\rm anc}](\bm x),
\end{equation}
where $\bm\Sigma_{\rm anc}=\sigma^2\bm B^\top[(2G-1)\bm I_2\oplus \bm I_{2(K-1)}]\bm B$ is the conditional covariance of the $K$ ancillary modes. This represents the fact that the amplified noise on the first mode is split amongst all the ancillary modes via $\bm B^\top$. The region ${\mathscr{I}_{\bm n}(\ell)\coloneqq\prod_{i=1}^{2K}\mathscr{I}_{n_i}(\ell)}$ is composed of intervals ${\mathscr{I}_{n_i}(\ell)\coloneqq[(n_i-1/2)\ell,(n_i+1/2)\ell]}$ indexed by integers $n_i\in\mathbb{Z}$. As an example, for $N=2$ modes, $f_{\bm 0}$ is the probability that the amplified ancillary displacement ${\bm x\sim\mathcal{N}(0,\sigma^2(2G-1)\bm I_2)}$ lies within the fundamental square region $\mathscr{I}_{\bm 0}=[-\ell/2,\ell/2]\times[-\ell/2,\ell/2]$.  

The qudit layer relies on the standard single-qudit ($d$-level), square GKP decoder~\cite{gkp2001}, which works as follows. An error syndrome, $\bm s_{\rm data}=\sqrt{d}\bm\Omega\bm\varepsilon \mod{\ell}$, is gathered from stabilizer measurements on the local data qudits following the O2O layer. Informed by the syndrome, a displacement, $\hat{D}({\Tilde{\bm\varepsilon}_{{\bm n}^\star}})$, is then applied, mapping the $q$ and $p$ quadratures of the local square lattices back to nearest lattice points, $\bm n^\star$. If the error, $\bm\varepsilon$, is too large, the error correction procedure will mistakenly map back to the wrong points, inducing a logical error. Consider a single qudit, and let $\bm\varepsilon=\varepsilon_q\oplus\varepsilon_p$. The error correction procedure will induce a logical $X$ error if the $q$-quadrature displacement, $\varepsilon_q$, is as a bit longer than half of the Pauli distance, $\mathfrak{D}_X=\sqrt{2\pi/d}$, i.e. when $\abs{\varepsilon_q}\gtrsim \sqrt{\frac{\pi}{2d}}$. The details involved in estimating error rates are ultimately governed by the PDF of the residual error, $P(\bm\varepsilon)$, from the O2O decoding layer. See Appendix~\ref{app_ssec:asymptotics} for more details.

\subsection{Example 1: Oscillators-to-oscillators}

\begin{figure}
    \centering
    \includegraphics[width=\linewidth]{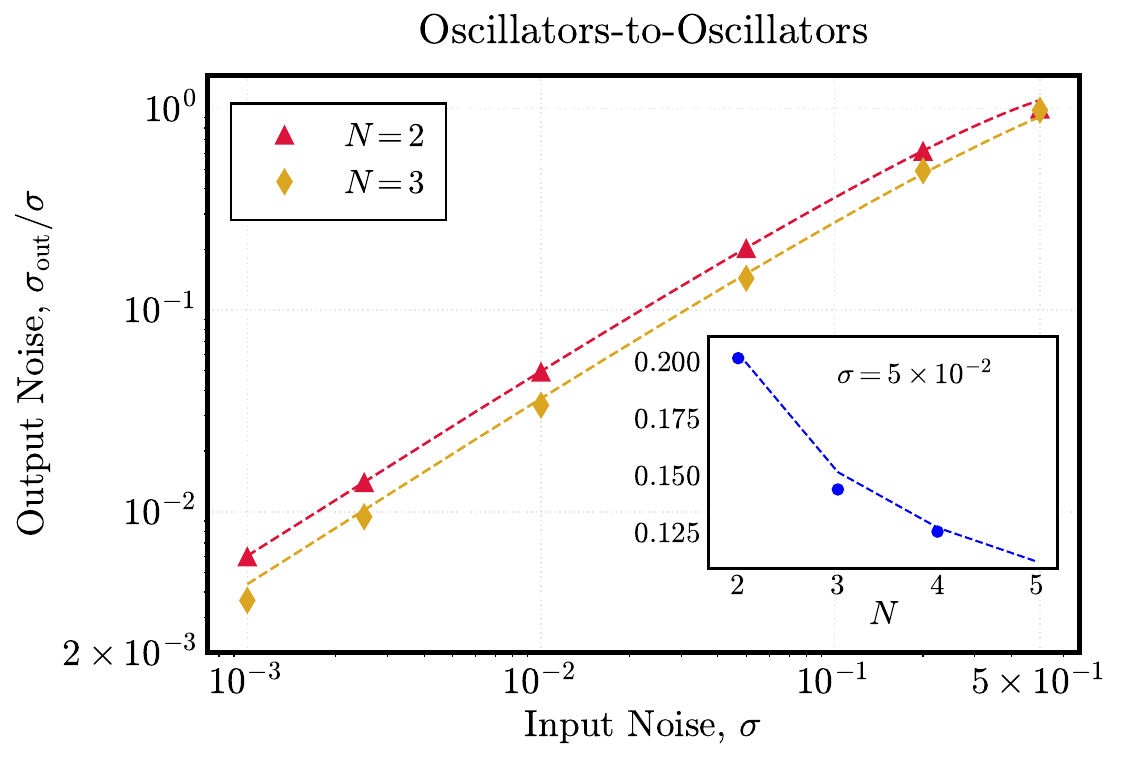}
    \caption{Output noise of dtms-O2O codes (Definition~\ref{def:dtms_o2o}) employing linear estimation. Results are normalized with respect to the input noise, i.e. $\sigma_{\rm out}/\sigma$. Analysis assumes uniform beamsplitting with no phases ($\phi=0$). Inset demonstrates scaling versus the number of modes $N$ for a few data points ($N=2,3,4)$. Dashed lines represent asymptotic expressions obtained from Eq.~\eqref{eq:sigma_out}.}
    \label{fig:o2o_sigmaout}
\end{figure}


The goal of an oscillators-to-oscillators code is to directly reduce the input additive noise, $\sigma$. Therefore, we evaluate the performance of the dtms-O2O code with the output variance, $\sigma^2_{\rm out}$ [generically defined in Eq.~\eqref{eq:output_variance}], of the code. We provide a heuristic argument that gives a good first-order estimate of $\sigma_{\rm out}$. Throughout, we assume a CSS-type configuration ($\phi=0$). 

As a first-order (Gaussian) approximation, $\sigma^2_{\rm out}\sim \Tr(\bm\Sigma_{\rm dtms})/2=\sigma^2/(2G-1)$, where $\bm\Sigma_{\rm dtms}$ is the conditional covariance given by Eq.~\eqref{eq:dtms_sigma} from O2O decoding. However, this approximation ignores discrete effects originating from the ancillary GKP modes, i.e.
\begin{equation}
    \sigma^2_{\rm out}=\frac{\sigma^2}{2G-1} ~+~ \text{``lattice effects''}.
\end{equation}
\QZ{ The ``lattice effects'' originate from uncertainty in estimating the noisy displacements from stabilizer measurements on the ancillary GKP modes. In particular, there is ambiguity of the actual size and direction of the noisy displacements due to the modularity of the GKP lattice. This modular uncertainty gets imprinted on the signal mode during the error corrective displacements in the O2O decoding.}
We can predict at what point these lattice effects become non-negligible, allowing us to estimate the scaling of the gain, $G$, with the input noise, $\sigma$, and consequently, the scaling of the output noise, $\sigma_{\rm out}$. \QZ{ For further mathematical details, please refer to Appendix~\ref{appssec:o2o}.}

Prior to stabilizer measurements, the single-mode displacement noise, $\bm x$, on any given ancillary GKP mode is Gaussian with variance $\sim2\sigma^2G/N$. Mathematically, this follows by taking the marginal of the ancillary covariance matrix $\bm\Sigma_{\rm anc}$ [written just below Eq.~\eqref{eq:f_n}] and assuming $G/K\gg1$, $\sigma\ll 1$, and $K=N-1\approx N$. Intuitively, this is due to distributing the amplified noise of the first ancilla mode uniformly among all ancillary modes. Lattice effects in the error correction procedure emerge when this noise is roughly the size of the lattice spacing, i.e. $\abs{\bm x}\sim \ell$, implying that $G\sim \ell^2 N/2\sigma^2$. We therefore anticipate that the output error scales, at best, as $\sigma^2_{\rm out}\sim \sigma^4/(\ell^2 N)$.

Through asymptotics (see Appendix~\ref{app_ssec:asymptotics}), we derive formulae for the optimal gain, $G_{\rm O2O}$, and resulting output variance, $\sigma_{\rm out}^2$, of the dtms-O2O code,
\begin{align}
    G_{\rm O2O}&\approx \frac{\pi (N-1)}{8\sigma^2}\left\{\ln\left(\frac{\pi^{3/2}(N-1)^2}{2\sigma^4}\right)\right\}^{-1},\label{eq:gain_o2o} \\
    \sigma_{\rm out}^2&\approx \frac{4\sigma^4}{\pi(N-1)}\ln\left(\frac{\pi^{3/2}(N-1)^2}{2\sigma^4}\right),\label{eq:sigma_out}
\end{align}
where $N\geq2$. This result is consistent with the heuristic scaling from above and aligns with the original $N=2$ two-mode squeezing code introduced in Ref.~\cite{noh2020o2o}. In Fig.~\ref{fig:o2o_sigmaout}, we plot results from a brute-force numerical optimization (data points). The asymptotic formulae (dashed lines) agree well with the numerical results. 

Observe that we obtain a quadratic suppression in the variance due to squeezing and a simultaneous suppression that is linear in the number of modes, $N$, due to the beamsplitters. We note that a comparable $1/N$ scaling was identified in Ref.~\cite{xu2022qubit} via augmented repetition codes. However, those codes do not exhibit the additional quadratic suppression in the variance from squeezing. 


\subsection{Example 2: Single-qudit codes}\label{ssec:ex2_qudit}

We apply the two-stage decoder to dtms single-qudit codes (Definition~\ref{def:dtms_qudit}) and infer logical $X$-type errors, corresponding to spurious displacements along the $q$ direction in phase space. We consider CSS-type codes here, so $X$ and $Z$ error are uncorrelated and of equal magnitude. [$Y$ errors are parametrically suppressed since $\mathfrak{D}_Y=\sqrt{2}\mathfrak{D}_X$.] 

One benefit of our simple two-stage decoder is that we can derive analytical expressions for the logical $X$ (or $Z$) error rate $\Pr[X]$ and infer an effective GKP code distance for arbitrary number of modes, $N$, and qudit dimension, $d$. We first present the qubit case ($d=2$) to gain some familiarity then generalize to arbitrary qudit dimension.


Following two-stage decoding, we find that the logical $X$ error rate is, to a good approximation, given by (see Appendix~\ref{app_ssec:asymptotics} for asymptotics),
\begin{equation}\label{eq:px_dtms}
    \Pr[X]\approx N\times {\rm erfc}\left(\sqrt{\frac{\mathfrak{D}_{\rm eff}^2(N)}{8\sigma^2}}\right),
\end{equation}
and the effective code distance $\mathfrak{D}_{\rm eff}(N)$ is defined as
\begin{equation}\label{eq:D_N}
    \mathfrak{D}_{\rm eff}(N)\coloneqq\sqrt{2G(N)-1}\mathfrak{D}_{\square}, 
\end{equation}
where $\mathfrak{D}_{\square}=\sqrt{\pi}$ and
\begin{equation}\label{eq:G_N}
    G(N)=1 + \frac{\sqrt{N^2+4N-4}-N}{4}
\end{equation}
is the optimal gain for linear decoding in the low-noise regime. The quantity $\mathfrak{D}_{\rm eff}(N)$ is the effective code distance deduced from the two-stage linear decoder and represents the ``Linear decoding bound'' plotted in Fig.~\ref{fig:code_distance}.

Unfortunately, the effective code distance $\mathfrak{D}_{\rm eff}(N)$ is bounded in the large $N$ limit,
\begin{equation}\label{eq:limit_DN}
    \lim_{N\rightarrow\infty}G(N)=3/2\implies \mathfrak{D}_{\rm eff}(N)\leq \sqrt{2}\mathfrak{D}_{\square}=\ell,
\end{equation}
which apparently originates from the simple linear decoder that we employ. This is highlighted by the fact that the results here contrasts the distances inferred from numerics (see Fig.~\ref{fig:code_distance}), which do not reference any particular decoder. This observation justifies that linear decoding---though simple and efficient---is not optimal, and there are limitations to its use. On the other hand, for the small code instance of $N=2$, the distance inferred from linear decoding matches the numerically optimized CSS dtms code (and Tesseract code) presented in Fig.~\ref{fig:code_distance}, suggesting that the two-stage decoder works quite well for the case of two modes.

\begin{figure}
    \centering
    \includegraphics[width=\linewidth]{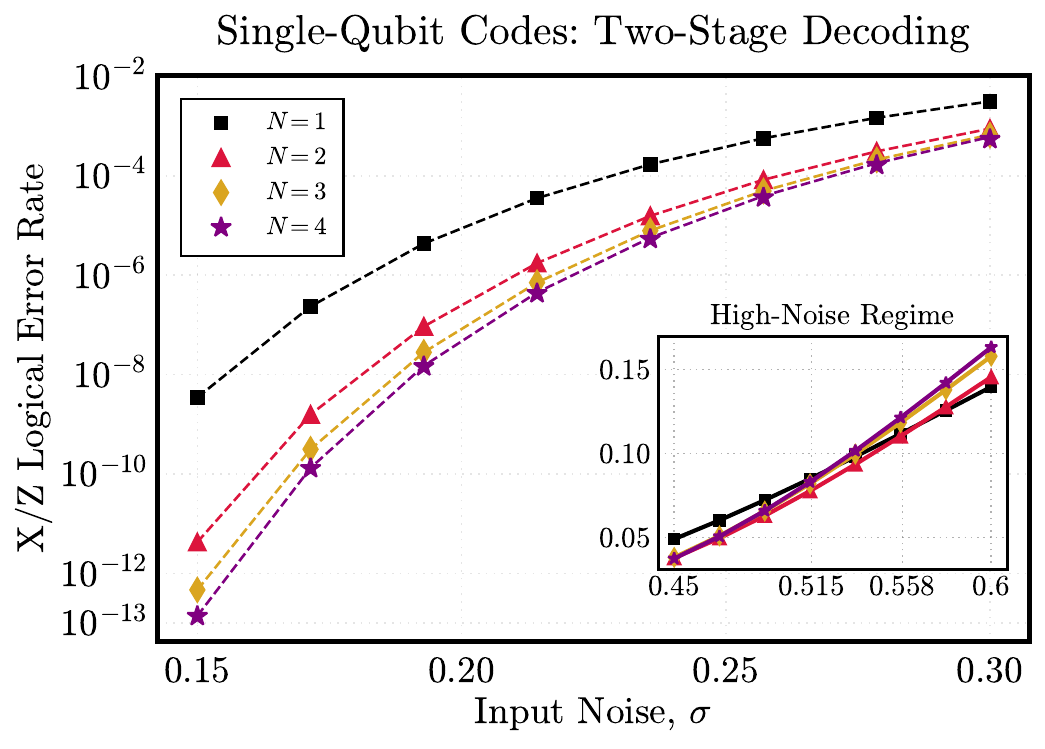}
    \caption{Logical $X$ (or $Z$) error rates for single-qubit dtms codes (Definition~\ref{def:dtms_qudit}) employing a two-stage linear decoder. Dashed lines represent asymptotic expressions from Eq.~\eqref{eq:px_dtms}. Inset highlights a notable crossing in the high-noise regime, indicative of threshold-like behavior near $\sigma\approx .558$.}
    \label{fig:logical_error}
\end{figure}

In Fig.~\ref{fig:logical_error}, we plot X (or Z) error rates for $N=1,2,3,4$ modes. Data points correspond to numerics while dashed lines represent the asymptotic formula~\eqref{eq:px_dtms}. For the numerical results, we perform numerical integration and numerically optimize the gain in the low-noise regime (at $\sigma=.15$). We find that the numerically optimized gain agrees well with the analytical result, $G(N)$, above. We then fix the gain to the low-noise value and subsequently compute the error rates for all other input noises, $\sigma$. This numerical approach agrees well with the asymptotics. 

In terms of error correction performance, we clearly observe an advantage for $N>1$ number of modes, however the relative advantage diminishes as $N$ increases, due to the limitations placed by linear decoding---specifically, a capped code distance. However, for small code sizes ($N\leq 4$), a notable advantage over single-mode encodings is found, spanning nearly 5 orders of magnitude for low noise ($\sigma\lesssim .15$). 

In the inset of Fig.~\ref{fig:logical_error}, we plot the error rate in the high-noise regime ($\sigma\gtrsim.4$). Notably, we observe a crossing for higher-mode encodings around $\sigma\approx .558$, indicating threshold-like behavior. This value aligns with previous results for qubit codes (see, e.g., Ref.~\cite{lin2023closest} and references therein) and, interestingly, also corresponds to the break-even point for O2O codes employing a linear decoder~\cite{noh2020o2o, wu2023optimal}. We speculate that, similar to these codes, the threshold can be extended (possibly to $\sigma= 1/\sqrt{e}\approx .607$) with better decoders~\cite{wu2023optimal} (see also Ref.~\cite{lin2023closest}).

We now extend the results to qudits ($d>2$). In gist, encoding higher dimensional systems allows us to marginally increase the decoding rate without sacrificing too much in the distance, as we now demonstrate. From asymptotics (see Appendix~\ref{app_ssec:asymptotics}), we infer an effective qudit code distance, 
\begin{equation}
    \mathfrak{D}_{\rm eff}(N,d)\coloneqq \sqrt{\frac{(2G(N;d)-1)2\pi}{d}},
\end{equation}
with gain,
\begin{equation}\label{eq:G_Nd}
    G(N;d)=1+\frac{\sqrt{(N-2)^2+4d(N-1)}-N}{4}.
\end{equation}
For $d=2$, the code distance and gain reduce to Eqs.~\eqref{eq:D_N} and~\eqref{eq:G_N}, respectively. For any fixed $d$, the gain is bounded, $\lim_{N\rightarrow\infty}G(N;d)=(1+d)/2$. Consequently, the code distance is also bounded, $\mathfrak{D}_{\rm eff}(N;d)\leq \ell$, analogous to the qubit case in Eq.~\eqref{eq:limit_DN}. 

One simple but concrete example is a two-mode qutrit code ($N=2$, $d=3$), exhibiting an effective distance $\mathfrak{D}_{\rm eff}(2,3)=\ell/3^{1/4}$. Interestingly, this is equal to the code distance of a (single-mode) GKP hexagonal qubit. Hence, dtms qutrit codes exist with code distance at least as large as the GKP hexagonal qubit code. Performance can be improved through better decoding or utilization of a non-CSS dtms code. [Note that the GKP hexagonal qubit code is non-CSS.]

The effective code distance for the two-stage decoder is bounded by a constant ($\leq \ell$), yet there is potential for a modest, $N$-dependent improvement in the encoding rate, $\log_2(d)/N$, beyond the single-qubit baseline of $1/N$. The following heuristic argument provides a rough scaling. For linear decoding, the optimal gain scales as $G\sim d$ for large $N$ and $d$. To implement decoding at the O2O level (first stage of the decoder) without inducing an error at the qudit level, the distributed noise on the ancillary modes, $G\sigma^2/N\sim d\sigma^2/N$, must satisfy $d\sigma^2/N\lesssim\ell^2$. If we increase $d$ arbitrarily, the code will be susceptible to arbitrarily small noise $\sigma$, unless $N$ increases proportionally. We thus establish that $d\lesssim \order{N}$, implying $\log_2(d)/N \lesssim \order{\log_2(N)/N}$.


\begin{figure}
    \centering
    \includegraphics[width=\linewidth]{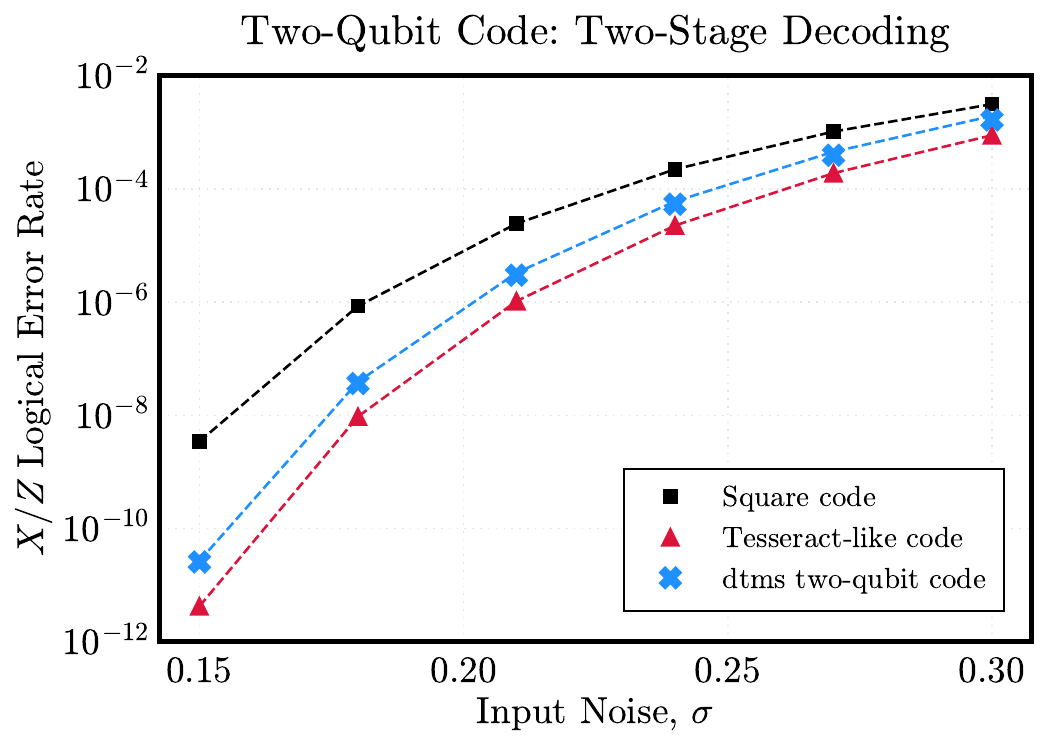}
    \caption{Single-qubit logical $X$ (or $Z$) error rate for  ($N=4$, $k=2$) dtms two-qubit code employing a two-stage linear decoder. Square code and Tesseract-like code are shown for comparison. Dashed lines represent asymptotic expressions from Eq.~\eqref{eq:px_dtms} for square ($N=1$) and Tesseract-like ($N=2$) codes, respectively, and Eq.~\eqref{eq:px_42} for the dtms two-qubit code.}
    \label{fig:px_42}
\end{figure}

\subsection{Example 3: Two-qubit codes}

We now apply the two-stage decoder to dtms two-qubit codes (Definition~\ref{def:dtms_2qubit}) and infer $X$-type errors. Again we consider CSS-type codes. The major difference here is that we are dealing with two qubits encoded into $N$ modes, thus both single-qubit and two-qubit errors are possible.

Employing asymptotics, as was done for single-qudit code, we can estimate an effective code distance for dtms two-qubit codes. With an optimized gain,
\begin{equation}\label{eq:2qubit_gain}
    G(N)=\frac{\left(\sqrt{4 N-7}+3\right)}{4},
\end{equation}
the effective code distance takes a simple form,
\begin{align}
    \mathfrak{D}_{\rm eff}^{(2)}(N)&=\sqrt{\frac{2G(N)-1}{G(N)}}\mathfrak{D}_{\square} \\
    &=\sqrt{\frac{\sqrt{4 N-7}+1}{\sqrt{4 N-7}+3}}\sqrt{2\pi}.
\end{align}
Similar to the single-qudit case, the distance is bounded by $\ell=\sqrt{2\pi}$ for large $N$. The effective distance has a different functional dependence on the gain, $G$, that differs from what is anticipated for the true distance derived from the logical dual matrix of the code~\eqref{eq:2qubit_distance} (see also numerical results in Section~\ref{ssec:numerics}). In particular, the effective distance is strictly smaller than the true distance, which we attribute to linear decoding. Nevertheless, we achieve notable performances with the proposed decoder. 

For instance, consider the $N=3$ dtms two-qubit code, which encodes two qubits into three modes using a 50:50 beamsplitter and a single two-mode squeezer. The effective code distance is $\mathfrak{D}_{\rm eff}^{(2)}(3)=\sqrt{\sqrt{5}-1}\mathfrak{D}_{\square}$, which deviates from our numerical findings (Section~\ref{ssec:numerics}), highlighting once more the general non-optimality of our two-stage decoder. Despite this, this code outperforms single-mode encodings. Specifically, the code distance, $\mathfrak{D}_{\rm eff}^{(2)}(3)$, is 1.11 times larger than the GKP square qubit code and 1.03 times larger than the GKP hexagonal qubit code.

We now focus on encoding two qubits into four modes, which requires two 50:50 beamsplitters and a single two-mode squeezer (see Fig.~\ref{fig:dtms42_encoder} for an illustration). The effective code distance is,
\begin{equation}
    \mathfrak{D}_{\rm eff}^{(2)}(4)=\sqrt{\frac{4\pi}{3}},
\end{equation}
which is, curiously, equal to the true distance of the smaller ($N=3$, $k=2$) code obtained from numerics (see Section~\ref{ssec:numerics}). The effective code distance, $\mathfrak{D}_{\rm eff}^{(2)}(4)$, is $\sqrt{4/3}\approx 1.15$ times larger than the square code. Furthermore, the quoted performance can be achieved with approximately 5.7 dB of squeezing ($G=3/2$).

We now consider error rates. Since we encode two qubits into multiple modes, there can generically exists correlations between $X_1$ and $X_2$ errors. We thus examine the likelihood of any error occurring in the joint space $X_1\cup X_2$, including both single-qubit and two-qubit $X$ errors. For codes with independent errors (a trivial example being two independent single-qubit codes), the joint error rate can be expressed as $\Pr[X_1\cup X_2]=\Pr[X_1]+\Pr[X_2]-\Pr[X_1]\Pr[X_2]$. The last term can be neglected to a good approximation when the noise is small, simplifying the joint error to the sum of single-qubit errors. 

From asymptotics, we derive analytical expressions for the error rates of dtms two-qubit codes ($N-2$ ancillae),
\begin{equation}\label{eq:px_42}
    \Pr[X_1]=\Pr[X_2]\approx (N-1)\times {\rm erfc}\left(\sqrt{\frac{\mathfrak{D}_{\rm eff}^2(N)}{8\sigma^2}}\right),
\end{equation}
and
\begin{equation}\label{eq:px_joint_42}
    \Pr[X_1\cup X_2]\approx N\times {\rm erfc}\left(\sqrt{\frac{\mathfrak{D}_{\rm eff}^2(N)}{8\sigma^2}}\right),
\end{equation}
analogous to previous results for dtms qudit codes~\eqref{eq:px_dtms}. Notably, for dtms two-qubit codes, we find that the likelihood of a two-qubit error is comparable to the likelihood of a single-qubit error, i.e. $\Pr[X_1\cap X_2]\propto \Pr[X]$. This is due to the structure of the code, which correlates the first and second mode by a 50:50 beamsplitter (see Fig.~\ref{fig:dtms42_encoder}). Hence, if an error happens on the first (second) qubit, an error likely happens on the second (first) qubit as well. The presence of correlated errors poses a potential drawback for the code. Although, if an error on one qubit can be flagged in some way, then we know with high probability that an error occurred on the other qubit and can account for that.

In Fig.~\ref{fig:px_42}, we plot the logical single-qubit error rate ($X_i$ or $Z_i$) for $N=4$ modes and compare the performance of the resulting dtms two-qubit code to four copies of a square qubit and two copies of a Tesseract-like qubit, respectively. Data points correspond to numerics while the dashed line for the dtms two-qubit code corresponds to the asymptotic formula~\eqref{eq:px_42}. Numerical optimization for the gain was performed in the low-noise regime, analogous to single-qudit codes described in the previous section, which agrees well with Eq.~\eqref{eq:2qubit_gain}.

Although using four independent square qubits yields a higher encoding rate ($k/N=1$), the error rate is correspondingly higher---around 3 or 4 orders of magnitude higher on average compared to the other encoding schemes. The Tesseract-like qubit (i.e., $N=2$ dtms qubit code) features a GKP code distance $\sqrt[4]{2}\sqrt{\pi}$, which is 1.03 times higher than the effective distance of the dtms two-qubit code (though numerical results indicate a larger true distance for the dtms code, Section~\ref{ssec:numerics}). However, two copies of a Tesseract-like qubit requires two squeezers to generate the code, whereas the dtms two-qubit code requires only one squeezer. This observation reflects a tradeoff between code performance and the physical resources (number of active elements) required to generate the code, at least when employing our two-stage decoder.


\section{Discussion} \label{sec:discussion}

In this work, we introduce a family of versatile distributed two-mode squeezing (dtms) GKP codes tailored for near-term (small) quantum information processors, which are well-suited for applications in quantum repeaters~\cite{azuma2023RMPQuInternet,fukui2021allOptical,rozpkedek2021quantum,wu2022continuous,schmidt2023RepeaterGKPqudits,rozpedek2023allphotonic} and quantum sensor networks~\cite{zhuang2020GKP_dqs}. Our proposed dtms GKP codes offer a unified framework capable of supporting both discrete and analog quantum information. The codes adopt a single two-mode squeezing element and simple passive interferometers for encoding, typically requiring squeezing levels below $10$dB. Our work furthermore spotlights the synergy between distributed quantum sensing protocols and the design of quantum error correction codes. Indeed, the syndrome measurement in error correction can be considered as a quantum sensing process. 

We summarize a few future directions that have emerged from our analyses. While our emphasis has been on simple two-stage decoders, which demonstrate good performance for small system sizes, finding better (yet efficient) decoders for larger systems is an outstanding open problem. Moreover, our focus on an iid noise model for all data and ancilla contrasts with real-world noise, which may be heterogeneous and potentially correlated. Addressing this problem calls for additional optimization in encoding and decoding steps, as explored for the oscillators-to-oscillators codes in Ref.~\cite{wu2021continuous}. Related open problems on the noise model also include the optimal correction for bosonic loss, where a commonly adopted (though suboptimal~\cite{zheng2024nearoptimalQEC}) approach is to convert loss to additive noise via pre-amplification.

\begin{figure}
    \centering
    \includegraphics[width=\linewidth]{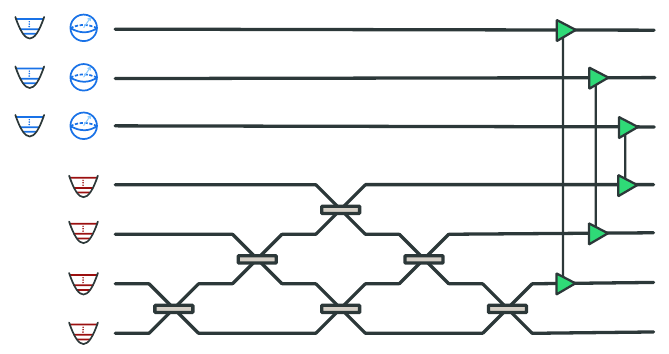}
    \caption{Introducing more squeezing components to encode more data modes. The number of squeezers scales with $k$ rather than $N$. Here, $N=7$ and $k=3$.}
    \label{fig:dtms_kmodes}
\end{figure}

We have focused on dtms GKP codes that rely on a single two-mode squeezing element, which seems sufficient for small-sized codes ($k=1,2$). \QZ{Besides the codes that we have focused on in this paper}, as the number of data modes increase, it becomes important to see how dtms codes can benefit from a few more squeezing elements. \QZ{ One option involves sparsely scattering two-mode squeezing operations amongst the large number of modes $N$ (such that the number of squeezing operations depends weakly on $k$ or $N$) and use beamsplitters to distribute the sparse squeezing throughout the entire network, in a similar vein to the dtms codes presented in this work. Further investigation is warranted here.} Another concrete option is to have a two-mode squeezing operation per data mode, thus allowing the number of active components to scale (linearly) with $k$ rather than directly with the size of the code block, $N$. In this scenario, one can leverage beamsplitter interference among GKP ancillae to benefit from multi-partite entanglement, similar to the dtms-GKP codes elaborated in this paper. See Fig.~\ref{fig:dtms_kmodes} for an illustrative schematic. It is worthwhile to draw parallels between these concepts and distributed quantum sensing of multiple parameters~\cite{zhuang2019physical}. In sensing context, the number of squeezed vacua increases with the number of parameters of interest, albeit with diminishing quantum advantage as more local parameters are estimated with a fixed number of modes. Likewise, as the number of data qudits, $k$, increases, a coding block of fixed size, $N$, provides diminishing protection of more logical information.


\begin{acknowledgements}
This project is supported by the Defense Advanced Research Projects Agency (DARPA) under Young Faculty Award (YFA) Grant No. N660012014029. QZ also acknowledges support from Office of Naval Research Grant No. N00014-23-1-2296, National Science Foundation OMA-2326746, 
National Science Foundation Grant No. 2330310, NSF CAREER Award CCF-2142882 and National Science Foundation (NSF) Engineering Research Center for Quantum Networks Grant No. 1941583.

QZ initiated and supervised the project. AJB proposed the distributed two-mode squeezing code, while JW proposed to add phase optimization to balance the code, in discussions with all authors. JW performed the numerical analyses, including initial stage of exploring code designs, solving the code distance and generated data for Fig. 3, under the supervision of AJB and QZ. JW wrote Appendix C and parts of Appendices A and B. AJB performed the asymptotic analyses and numerical analyses to obtain Fig. 5, Fig. 6 and Fig. 7. AJB wrote the major part of the manuscript, plotted all figures, with inputs from QZ and JW. 

\end{acknowledgements}

\appendix

\section{Theory of multi-mode GKP Codes}\label{app:theory_gkp}

To better elucidate the findings presented in this paper, we provide a review of multi-mode GKP states, including some basic elements of classical lattice theory. With the high-level lattice technology in hand, we leap to the quantum theory of multi-mode GKP states through the stabilizer formalism. Technical details omitted here can be found in Refs.~\cite{conway2013sphere,gkp2001,baptiste2022multiGKP, eisert2022lattice,brady2023GKPrvw}. Lastly, we highlight key ingredients for characterizing GKP qudit codes, such as the GKP code distance, and GKP oscillators-to-oscillators codes, such as analog noise suppression via two-mode squeezing.


\subsection{Stabilizer formalism for GKP states}


We associate an $N$-mode GKP state $\ket{\Psi_{\mathcal{L}}}$ with a $2N$-dimensional symplectically integral lattice $\mathcal{L}$ in the phase space, $\mathbb{R}^{2N}$, where each position ($q$) and momentum ($p$) quadrature constitutes a dimension. The lattice $\mathcal{L}$ can be described by a set of $2N$-dimensional basis vectors $\{\bm m_j\}_{j=1}^{2N}$, which we package into a generator matrix,
\begin{equation}
    \bm M \coloneqq (\bm m_1, \bm m_2, \dots, \bm m_{2N}).
\end{equation}
The lattice is symplectically integral in the sense that the symplectic product between any two basis vectors is an integer (i.e., $\bm m_i^\top\bm\Omega\bm m_j\in\mathbb{Z}$). We codify this property compactly in the symplectic Gram matrix,
\begin{equation}
    \bm A\coloneqq\bm M^\top \bm\Omega\bm M,
\end{equation}
such that $\bm A_{ij}\in\mathbb{Z}$.

Starting from any point $q\in\mathcal{L}$, we can reach any other point in $\mathcal{L}$ via discrete displacements along the vectors $\bm m_j$ (columns of $\bm M$). In this sense, $\bm M$ generates the lattice, and we can use this interpretation to define the lattice, $\mathcal{L}\coloneqq\{\bm M^\top\bm a\,|\,\bm a\in\mathbb{Z}^{2N}\}$. The representation $\bm M$ of $\mathcal{L}$ is not unique because we are free to multiply $\bm M$ by an unimodular matrix $\bm N$ (i.e., $\bm N_{ij}\in\mathbb{Z}$ and $\det\bm N=\pm 1$), such that $\bm M^\prime=\bm M\bm N$ and $\bm M$ are both faithful representations of $\mathcal{L}$. Clearly, this change of basis does not alter the definition of $\mathcal{L}$ given above but is otherwise useful to examine properties of the lattice.
\begin{theorem}[Appendix B of Ref.~\cite{lin2023closest}]\label{thm:normal_form}
For any anti-symmetric matrix $\bA$ with integer elements, there exists a unimodular matrix $\bN$ such that
\begin{align}
    \bN^\top \bA \bN = \bm\Omega\left(\bigoplus_{j=1}^N d_j\bm I_2\right)
    \label{eq:block-diagonal-A}
\end{align}
where $d_i \in \Integer^+$.
\end{theorem}
As we elaborate more below, the positive integer $d_j$ here corresponds to a $d_j$-level system (a qudit) encoded in the $j$-th mode of a $N$-mode bosonic system.

Along with the lattice $\mathcal{L}$ we have the dual lattice $\mathcal{L}^{\perp}$ which consists of all vectors $\bm v$ that have integral symplectic product with the vectors in $\mathcal{L}$, such that ${\mathcal{L}^{\perp}\coloneqq\{\bm v|\bm m^\top\bm \Omega\bm v\in\mathbb{Z}\,\forall\,\bm m\in\mathcal{L}\}}$. It follows that $\mathcal{L}\subseteq\mathcal{L}^{\perp}$. We can choose a generator basis for the dual lattice as
\begin{equation}\label{eq:dual_M}
    \overline{\bm M} \coloneqq \bm M \bm A^{-1}\bm \Omega,
\end{equation}
such that $\bm M^{\top}\bm\Omega\overline{\bm M}=\bm\Omega$.

The quotient space $\mathcal{L}^{\perp}/\mathcal{L}$ is the space in which we can encode discrete logical information (e.g., qudits). Indeed, $\dim(\mathcal{L}^{\perp}/\mathcal{L})=\det\bm M/\det\overline{\bm M}=\prod_{j=1}^N d_j^2$ provides the number of logical operations in the code space~\cite{gkp2001,baptiste2022multiGKP,eisert2022lattice}, with each $d_j$ describing the number of discrete states encoded into the $j$-th subsystem (or mode).
\begin{definition}[Subsystem codes]\label{def:subystem_codes}
 We refer to each $d_j$ as the local code dimension of the $j$-th subsystem (or mode). The subsystem encodes $m$ (qu)bits if $d_j = 2^m$. We refer to the subsystem as canonical or self-dual if $d_j = 1$.
\end{definition}

Theorem~\ref{thm:normal_form} suggests that $\bm M\propto (\bigoplus_{j=1}^N\sqrt{d_j}\bm I_2)$, up to a scaling matrix that leaves the symplectic form $\bm\Omega$ unchanged. This intuition happens to be correct.
\begin{theorem}[Codes and symplectic transforms]\label{thm:codes_and_S}
    Consider a lattice $\mathcal{L}$ represented by $\bM$, with subsystems $d_j$. Then, up to a unimodular transformation, the generator matrix can be written as
    \begin{align}
        \bM = \bm S\left(\bigoplus_{j=1}^N\sqrt{d_j}\bm I_2\right),
    \end{align}
    where $\bS\in{\rm Sp}(2N,\mathbb{R})$.
\end{theorem}
See Ref.~\cite{eisert2022lattice} for details (Corollary 1). An interesting implication of Theorem~\ref{thm:codes_and_S} is that the generator matrix for any $[[N,k]]$ code, which encodes $k$ (qu)bits into $N$ modes, can be expressed as $\bm M_{[[N,k]]}=\bm S_{[[N,k]]}(\sqrt{2}\bm I_{2k}\oplus\bm I_{2(N-k)})$. We illustrate this with examples in Appendix~\ref{app:example_codes}. This also affords an operational interpretation for code construction in the quantum theory, as $\bm S$ can be linked to a Gaussian unitary (see below for further discussion).

It is often useful to consider special classes of codes that have CSS-type properties, such that stabilizer checks decompose into $X$- and $Z$-type stabilizers. Following Ref.~\cite{eisert2022lattice}, we generalize the CSS property to lattice codes.
\begin{definition}[CSS property~\cite{eisert2022lattice}]
    A lattice (e.g., GKP) code is said to be a CSS lattice code if its stabilizers can be decomposed into $q$ and $p$ stabilizers. In other words, the generator matrix for a CSS GKP code, upon reordering of the $q$ and $p$ rows/columns, can be written as
    \begin{equation}
        \bm M_{\rm CSS}= \bm M_q\oplus\bm M_p.
    \end{equation}
\end{definition}
Simple examples of CSS GKP codes are square and rectangular GKP qubits. Concatenating a local CSS GKP code (e.g., square code) with an outer CSS qubit code maintains the CSS property of the code~\cite{gkp2001,eisert2022lattice}. From Theorem~\ref{thm:codes_and_S}, we can also generate a CSS GKP code by acting on square qudits with a symplectic transformation (Gaussian unitary) that separates into $q$ and $p$ blocks.

We now leap to the quantum theory of GKP states through the stabilizer formalism applied to bosonic modes~\cite{gottesman1997stabilizer,gkp2001}. A $2N$-dimensional stabilizer group $\mathcal{G}$ is an abelian group generated by a set of $2N$ commuting operators, $\hat{G}_j$, which act on the $N$-mode bosonic Hilbert space $\mathscr{H}$. We define the group abstractly through the generators as $\mathcal{G}\coloneqq\langle\hat{G}_1, \hat{G}_2, \dots, \hat{G}_{2N}\rangle$. A stabilizer code space $\mathcal{C}\subset\mathscr{H}$ associated with $\mathcal{S}$ is then defined as the $+1$ eigenspace of $\mathcal{G}$, i.e. ${\mathcal{C}\coloneqq\{\ket{\Psi} \,\big|\, \hat{G}\ket{\Psi}=\ket{\Psi},\,\forall\, \hat{G}\in\mathcal{G}\}}$. 

Combining the stabilizer formalism with lattice theory, we define a GKP stabilizer group, from which we construct a GKP (stabilizer) code~\cite{gkp2001}. 

\begin{definition}[GKP stabilizer group]
    Consider a $2N$-dimensional lattice $\mathcal{L}$ with representation $\bm M$. We define stabilizers as displacements along the vectors $\bm m_j$ (columns of $\bm M$), such that
    \begin{equation}
        \hat{S}(\bm m_j)\coloneqq \hat{D}(\ell\bm m_j),
    \end{equation}
    where $\hat{D}(\cdot)$ is the displacement operator in Eq.~\eqref{eq:displacement}. For brevity, we denote stabilizers as $\hat{S}_j$ when the context is clear. The GKP stabilizer group associated with the representation $\bm M$ of $\mathcal{L}$ is then,
    \begin{equation}
        \mathcal{S}(\bm M)\coloneqq\langle \hat{S}_1, \hat{S}_2,\dots,\hat{S}_{2N}\rangle.
    \end{equation}
\end{definition}

Due to the integral conditions imposed on the vectors $\bm{m}_j$ (i.e., $\bm{m}_i^\top\bm{\Omega}\bm{m}_j\in\mathbb{Z}$) and the commutation relation for displacement operators [see Eq.~\eqref{eq:displacement_comm}], it is evident that the operators $\hat{S}_j$ (and their products) commute. Using this set of stabilizer generators, we construct a GKP code.
\begin{definition}[GKP code~\cite{gkp2001}]
    A GKP code space $\calC(\bm M)$ is defined as the +1 eigenspace of the GKP stabilier group $\mathcal{S}(\bm M)$, i.e.
    \begin{equation}
        \calC(\bm M)\coloneqq \Big\{\ket{\Psi_{\mathcal{L}}} \,\Big|\, \hat{S}\ket{\Psi_{\mathcal{L}}}=\ket{\Psi_{\mathcal{L}}},\,\forall\, \hat{S}\in\mathcal{S}(\bm M)\Big\},
    \end{equation}
    where $\bm M$ is a generator matrix for the lattice $\mathcal{L}$.
\end{definition}


To provide a more concrete perspective, we adopt an operational approach that can be employed to characterize (or create) a $[[N,k]]$ GKP qudit code. We assume a device that generates $k$ local (i.e., single-mode) square GKP qudits ($d_j>1$) and another device that generates a block of $N-k$ local, square canonical ($d=1$) GKP states. The initial generator matrix is a trivial direct sum,
\begin{equation}\label{eq:M_initial}
    \bm M_{\rm in}\coloneqq \left(\bigoplus_{j=1}^k\sqrt{d_j}\bm I_2\right)\oplus \bm I_{2(N-k)}.
\end{equation}
We then push this collection of $N$ independent GKP states through a multi-mode network comprised of active and passive elements, described by a Gaussian unitary $\hat{U}_{\bm S}$. This construction generates a multi-mode GKP qudit code with a representation,
\begin{equation}\label{eq:M_enc}
    \bm M_{L}\coloneqq\bm S\bm M_{\rm in},
\end{equation}
where $\bm S$ is the symplectic matrix corresponding to $\hat{U}_{\bm S}$. 

The stabilizers and code space of the resulting code, $\mathcal{S}(\bm M_{L})$ and $\mathcal{C}(\bm M_{L})$, respectively, are connected to the initial stabilizers $\mathcal{S}(\bm M_{\rm in})$ and code space $\mathcal{C}(\bm M_{\rm in})$ by unitary conjugation, yielding the simple correspondence
\begin{align}
    \hat{S}(\bm m_j)&\rightarrow \hat{S}(\bm S\bm m_j) \in\mathcal{S}(\bm M_{L})  \\ 
    \qq{and} \ket{\Psi_{\rm in}}&\rightarrow\hat{U}_{\bm S}\ket{\Psi_{\rm in}}\in\mathcal{C}(\bm M_{L})
\end{align}
where $\hat{S}(\bm m_j)\in\mathcal{S}(\bm M_{\rm in})$ and $\ket{\Psi_{\rm in}}\in\mathcal{C}(\bm M_{\rm in})$. 

Pauli $X$ and $Z$ operations (qudit flips and phase flips) acting on the initial code space $\mathcal{C}(\bm M_{\rm in})$ are related to the first $2k$ columns vectors ($\bar{\bm m}_{{\rm in},1}, \bar{\bm m}_{{\rm in},2}, \dots, \bar{\bm m}_{{\rm in},2k}$) of the dual matrix, $\overline{\bm M}_{\rm in}$ [see Eq.~\eqref{eq:dual_M} for its definition], where
\begin{equation}
\overline{\bm M}_{\rm in}=\bm M_{\rm in}^{-1}=\left(\bigoplus_{j=1}^k\frac{\bm I_2}{\sqrt{d_j}}\right)\oplus \bm I_{2(N-k)}.
\end{equation}
Whence, the Paulis on $\mathcal{C}(\bm M_{\rm in})$ correspond to single-mode displacements, written abstractly as ${\hat{X}_{{\rm in},j}=\hat{D}(\ell\bar{\bm m}_{{\rm in},2j-1})}$ and $\hat{Z}_{{\rm in},j}=\hat{D}(\ell\bar{\bm m}_{{\rm in},2j})$
for $j=1,2,\dots, k$ [see Eq.~\eqref{eq:qudit_paulis}]. 
We can connect the Paulis acting on the $k$-mode code space $\mathcal{C}(\bm M_{\rm in})$ to logical Paulis acting on the $N$-mode code space $\mathcal{C}(\bm M_{L})$ through the encoding $\bm S$. Given $\bm M_{L}$ in Eq.~\eqref{eq:M_enc}, the dual matrix after encoding takes a simple form,
\begin{equation}\label{eq:dual_ML}
    \overline{\bm M}_{L}=\bm S \bm M_{\rm in}^{-1},
\end{equation}
such that 
\begin{equation}
    \hat{X}_{L,j}=\hat{D}(\ell\bm S\bar{\bm m}_{{\rm in},2j-1}), \quad \hat{Z}_{L,j}=\hat{D}(\ell\bm S\bar{\bm m}_{{\rm in},2j})
\end{equation}
for $j=1, 2,\dots, k$. These are valid unitary representations of logical $X$ and $Z$ operations acting on the $N$-mode code space $\mathcal{C}(\bm M_{L})$. Although, they do not generically correspond to minimal-weight Paulis; see Definition~\ref{def:pauli_distance} and the surrounding discussions.

This concrete framework for code construction (via $\bm M_{\rm in}$ and $\hat{U}_{\bm S}$) not only serves illustrative purposes but potentially offers a practical approach to realize GKP codes in platforms relying on itinerant modes, such as optical quantum information processors. Though, this rests on the assumption that the multi-mode Gaussian unitary $\hat{U}_{\bm S}$, or some version thereof, can be effectively engineered. In this paper, we achieve this through a single two-mode squeezer and multi-port interferometers (see Fig.~\ref{fig:dtms_schematic} of the main text for an illustration).

\subsection{Qubits-to-oscillators: GKP code distance}\label{ssec:q2o}

When encoding discrete information into a multi-mode GKP state, we need some generic method of characterizing the performance of the resulting code. One natural figure of merit is the so-called \textit{GKP code distance}~\cite{gkp2001,baptiste2022multiGKP,eisert2022lattice}, which gives a relative measure of how large a displacement error needs to be in order to enact a logical error on the code space. For simplicity, we focus on a single-qubit code. The extension to qudits is straightforward.
\begin{definition}[GKP Pauli distance~\cite{lin2023closest}]\label{def:pauli_distance}
    Consider a GKP qubit code $\mathcal{C}(\bm M)\sim\mathbb{C}^2$ associated to a $2N$-dimensional lattice $\mathcal{L}$ with a representation $\bm M$. We define the GKP Pauli distance for the logical operator $\hat{J}_L\in\{\hat{X}_L,\hat{Y}_L,\hat{Z}_L\}$ as
    \begin{equation}
        \label{eq:Pauli-distance}\mathfrak{D}_J\coloneqq \ell\times\min_{\bm a\in\mathbb{Z}^{2N}}{\norm{\bar{\bm m}_J - \bm M \bm a}},
    \end{equation}
    where ${\bar{\bm m}_J\in \calL^{\perp}/\calL}$ is a Pauli displacement vector corresponding to the logical Pauli operation $\hat{J}_L$ and $\ell=\sqrt{2\pi}$.
\end{definition}
In the settings of interest here, the Pauli $X$ and $Z$ displacement vectors, $\bar{\bm m}_X$ and $\bar{\bm m}_Z$, respectively, correspond to the first two columns of the dual matrix $\overline{\bm M}$. This correspondence follows from the code construction in Eqs.~\eqref{eq:M_enc} and~\eqref{eq:dual_ML}. For example, a square qubit has $\bar{\bm m}_X=2^{-1/2}(1,0)$ and $\bar{\bm m}_Z=2^{-1/2}(0, 1)$, and thus $\mathfrak{D}_X(\square)=\mathfrak{D}_Z(\square)=\sqrt{\pi}$, and $\mathfrak{D}_Y(\square)=\sqrt{2\pi}$. It is important to note that the vectors $\bar{\bm m}_X$ and $\bar{\bm m}_Z$ of $\overline{\bm M}$ do not necessarily represent the shortest displacements capable of implementing a logical Pauli, hence the minimization in the definition above.

\begin{definition}[GKP code distance]
    Given the Pauli distances $\mathfrak{D}_J$ for a GKP qubit code $\mathcal{C}(\bm M)\sim\mathbb{C}^2$, the GKP code distance $\mathfrak{D}$ is defined as
    \begin{equation}
        \mathfrak{D}\coloneqq \min_{J\in\{X,Y,Z\}}\mathfrak{D}_J.
    \end{equation}
\end{definition}
The GKP code distance sets a length scale that serves as an initial estimate for assessing GKP code performance. In particular, given an error $\bm e\sim\mathcal{N}(0,\sigma^2\bm I_{2N})$, the displacement is correctable if $\sigma\lesssim \mathfrak{D}$.

Finally, we note that concatenating a local GKP code (of GKP distance $\mathfrak{D}_{\rm local}$) with a $[[N,k,d]]$ qubit code results in an increased GKP distance, $\mathfrak{D}_{\rm conc}\geq \sqrt{d}\mathfrak{D}_{\rm local}$. Equality is met for CSS-type codes~\cite{eisert2022lattice}. Examples include square GKP-surface codes~\cite{terhal2019toricGKP,noh2020surface}, with $d_{\rm surf}\sim\sqrt{N}\implies\mathfrak{D}_{\rm surf}\sim \sqrt{\pi}N^{1/4}$.



\subsection{Oscillators-to-oscillators: Reducing the variance}\label{ssec:o2o}

Protecting continuous-variable (i.e., analog) data against analog noises, such as additive noise or attenuation, is a distinct challenge that has only recently been addressed via GKP oscillators-to-oscillators (O2O) codes~\cite{noh2020o2o}.\footnote{Indeed, the original Gottesman-Kitaev-Preskill paper~\cite{gkp2001} stated that protecting analog information ``might be too much to hope for''.} The goal of O2O codes is to directly mitigate analog errors. That is, assuming some small noise $\sigma\ll 1$, we want $\sigma\rightarrow \sigma^\alpha$ for some $\alpha>1$. This capability has applications in distributed quantum sensing~\cite{zhuang2020GKP_dqs,zhou2022enhancing,Zhang2021dqs} (e.g., protecting a bright squeezed beam) and other continuous-variable information processing tasks.

Consider a multi-mode GKP O2O code that encodes $k$ oscillators into $N\geq k$ oscillators, using an encoding block of $N-k$ canonical GKP ancillae, and protects against the AGN channel $\Phi_\sigma$. Given the analog data state is $\Psi$, the output state from the GKP O2O decoding map is given by $\Phi_{P}(\Psi)$, where $\Phi_{P}$ is a non-Gaussian additive noise channel in Eq.~\eqref{eq:angn_channel} described by the PDF $P(\bm e)$. Here, $\bm e$ is the residual displacement noise on the analog data, such that $\expval{\bm e}=0$ and $(\bm V_{\rm out})_{ij}\coloneqq\expval{\bm e_i\bm e_j}\geq0$. The matrix $\bm V_{\rm out}$ is the output covariance matrix derived from the O2O code that characterizes the residual noise affecting the data. We use it as a performance metric for an O2O code via the output variance, $\sigma_{\rm out}^2\coloneqq\Tr{\bm V_{\rm out}}/2k$, as defined in Eq.~\eqref{eq:output_variance} of the main text. Given that we aim to safeguard an arbitrary analog state, the output noise variance of the code is the preferred metric, contrasting, for instance, the likelihood of a discrete Pauli error in DV codes.

The non-Gaussian PDF $P(\bm e)$ is independent of $\Psi$ but depends on the Gaussian encoder as well as the ancillae consumed, whether they be canonical square GKP, hexagonal GKP, etc. Hence, to compute the output variance explicitly, we must specify details of the encoding and decoding. On the other hand, some generic things can be said about the output noise of GKP O2O codes.

For iid AGN, it can be shown, using quantum capacity arguments for GKP codes~\cite{noh2020o2o,wu2023optimal}, that the output error is bounded by,
\begin{equation}\label{eq:capacity_bound}
        \sigma_{\rm out}^2 \geq \frac{1}{\sqrt{e}}\left(\frac{\sigma^{2}}{1-\sigma^{2}}\right)^{N/k}.
\end{equation}
Furthermore, for generic (not necessarily GKP) O2O codes that utilize unitary Gaussian operations for encoding and decoding, one can employ an effective two-mode squeezing decomposition (through the modewise entanglement theorem~\cite{reznik2003modewise}) to show that
\begin{equation}\label{eq:threshold_bound}
\sigma_{\rm out}^2\geq\frac{1}{k}\sum_{i=1}^{k}\frac{ \sigma^2}{2G_i-1},
\end{equation}
where $G_i$ are the gains of the TMS operations in the decomposition~\cite{wu2023optimal}. We thus see that, without some amount of squeezing, analog noise reduction is generally impossible. 

The squeezing bound~\eqref{eq:threshold_bound} implies that the noise can be arbitrarily suppressed, however this seems in contradiction with the lower bound for GKP codes~\eqref{eq:capacity_bound}. This paradox can be resolved by noting that the squeezing cannot be arbitrarily large in GKP codes. Specifically, if $\xi\sim\mathcal{N}(0,\sigma^2)$ is the original AGN, then the amplified error, after TMS operations, is $\xi_G\sim\mathcal{N}(0,(2G-1)\sigma^2)$. To unambiguously resolve the displacement error from syndrome measurements on the GKP ancilla, we must have that $\xi_G\lesssim\sqrt{\pi}/2\implies G\lesssim \pi/8\sigma^2+1/2$, otherwise harmful lattice effects emerge in the decoding process. Substituting this condition into Eq.~\eqref{eq:threshold_bound} (and assuming $N=M=1$ for simplicity), we find that $\sigma_{\rm out}^2\gtrsim 4\sigma^4/\pi$, in agreement with the scaling implied by Eq.~\eqref{eq:capacity_bound}. This crude estimate is actually quite close to a more precise accounting [see Eq.~\eqref{eq:sigma_out} in the main text].

\section{Details of dtms Codes}\label{app:DTMS_detail}


\subsection{Uniform beamsplitter array}
\label{app:beamsplitter}

Our dtms-GKP code design relies on a staircase of beamsplitters that uniformly distribute locally amplified noises to all modes [see Fig.~\ref{fig:dtms_schematic} and Eq.~\eqref{eq:dtms_encoder}]. We elaborate the choice of beamsplitter transmissivities in order to achieve uniform splitting. We write the symplectic matrix, $\bm B$, that describes the configurable multi-port interferometer in block form, 
\begin{equation}\label{eq:B_block}
    \bB = \begin{pmatrix}
        \Tilde{\bB} & \\
        & \bB^\prime
    \end{pmatrix},
\end{equation}
where $\Tilde{\bB}$ is a $2(N-1)\times2$ sub-matrix of $\bB$. When there are no phases ($\phi_j=0$), we can write $\Tilde{\bB}$ in block form,
\begin{align}
\Tilde{\bB}^\top &= \left(
    \pm \sqrt{\eta_1}\bI_2, \pm \sqrt{(1-\eta_1)\eta_2}\bI_2, \cdots,\right. \nonumber \\
    & \left. \qquad \pm \sqrt{(1-\eta_1)...(1-\eta_{N-2})}\bI_2 \right),
\end{align}
where $\eta_i \equiv |\cos{\theta_i}|^2$ is the $i$-th beamsplitter transmissivity. A balanced beamsplitter can be obtained by taking $\eta_1=(1-\eta_1)\eta_2=...=(1-\eta_1)...(1-\eta_{N-2})$. Solving this recursively, $\eta_{N-2}=1/2$, ..., $\eta_1=1/(N-1)$, leading to, 
\begin{equation}
\Tilde{\bB}^\top = \Big(
    \pm \bI_2, ..., \pm \bI_2\Big)/\sqrt{N-1}.
\end{equation}
Adding phases leads to the block form, 
\begin{equation}
\Tilde{\bB}^\top = \big(
     \bR(\phi_1),...,\bR(\phi_{N-1}) \big)/\sqrt{N-1},
\end{equation}
where $\bR(\phi_i)$ is the single-mode phase rotation in Eq.~\eqref{eq:phase}, and signs have been absorbed.


\subsection{Correlated noise matrix}\label{app:noise_matrix}
After the unitary dtms encoding-decoding stage, the initially independent data and ancillae noises, $\bm e\oplus \bm e_{\rm anc}\sim\mathcal{N}(0,\sigma^2\bm I_{2N})$, become correlated per ${\bm e^\prime\oplus \bm e_{\rm anc}^\prime=\bm S_{\rm dtms}^{-1}(\bm e\oplus \bm e_{\rm anc})\sim\mathcal{N}(0,\bm V)}$, where $\bm V=\sigma^2 \bS^{-1}_{\rm dtms}\bS^{-\top}_{\rm dtms}$ and $\bm S_{\rm dtms}=\left(\bm S_{G}\oplus\bm I_{2(N-2)}\right)\left(\bm I_{2}\oplus\bm B\right)$ [Eqs.~\eqref{eq:dtms_encoder} and~\eqref{eq:dtms_corrV}]. Writing the covariance out in the sub blocks per Eq.~\eqref{eq:B_block}, we have,
\begin{widetext}
\begin{equation}
    \bV = 
    \sigma^2\begin{pmatrix}
        (2G-1)\bI_2 & & -2\sqrt{G(G-1)}\bZ \Tilde{\bB} \\
        -2\sqrt{G(G-1)}\Tilde{\bB}^\top \bZ & & \bm B^\top\left((2G-1)\bm I_2\oplus \bm I_{2(N-2)}\right)\bm B
    \end{pmatrix}.
\end{equation}
\end{widetext}
The upper left block represents the locally amplified data noise. Observe that the ancillary noise can be interpreted as a single-mode thermal state (of $G-1$ photons) distributed to a collection of $N-1$ oscillators via $\bm B^\top$. For a given realization, $\bm e\oplus \bm e_{\rm anc}$, the correlated noises are,
\begin{align}
    \bm e^\prime &= \sqrt{G}\bm e -\sqrt{G-1}\bm Z\bm e_{{\rm anc},1}\label{eq:data_eprime}\\
    \qq{and} 
    \bm e_{\rm anc}^\prime &= \bm B^\top
    \begin{pmatrix}
        \sqrt{G}\bm e_{{\rm anc},1} -\sqrt{G-1}\bm Z\bm e \\
        \bm e_{\rm anc,2} \label{eq:anc_eprime}\\
        \vdots 
    \end{pmatrix}.
\end{align}
Thus, in principle, through correlations (off-diagonal elements of $\bm V$) and measurements of ancillary noises (lower right block of $\bm V$), the data noise can be almost exactly inferred for large $G$. There will be limits to $G$, however, due to the measurement process, as we detail in the following subsections. 

\subsection{Decoding dtms codes: Asymptotics}\label{app_ssec:asymptotics}

We sketch the derivations leading to the analytical results for our two-stage decoder (Section~\ref{sec:decoding_dtms}). We start with the oscillators-to-oscillators stage of decoding. See Fig.~\ref{fig:linear_decoder} in the main text for an illustrative reference. Then we derive asymptotic expressions for the output variance of oscillators-to-oscillators codes. Finally, we outline the argument leading to analytical expressions for logical error rates for qudit codes.

\subsubsection{O2O layer decoding: Error PDF}
Starting with the ancillary  noises in Eq.~\eqref{eq:data_eprime}, which are correlated with the data noises, we can perform stabilizer measurements on the ancillary GKP, obtaining an error syndrome, $\bm s_{\rm anc}=\bm\Omega\bm e_{\rm anc}^\prime \mod{\ell}$. Provided with these syndromes, we construct an estimator, $\Tilde{\bm e}^\prime$ for the data noises via linear estimation $\Tilde{\bm e}^\prime=\bm F_{\rm dtms}\bm\Omega^\top\bm s_{\rm anc}$,
where $\bm F_{\rm dtms}$ is a $2\times 2K$ rectangular matrix. For CSS-type dtms qudit and O2O codes ($\phi=0$), we choose
\begin{equation}
    \bm F_{\rm dtms}=\frac{-C_G}{\sqrt{K}}\left(\bm Z\, \bm Z\, \dots\, \bm Z\right)\eqqcolon \bm F,
\end{equation}
 where we drop the dtms subscript in this appendix for notational convenience. Here,
\begin{align}
C_G=2\sqrt{G(G-1)}/(2G-1),
\label{c_define}
\end{align}
which is also defined in Eq.~\eqref{eq:c_constant} of the main text.
This choice for the estimator minimizes the data noise variance in the Gaussian approximation. To see this, consider the multivariate Gaussian PDF, $g[\bm V](\bm e^\prime,\bm e_{\rm anc}^\prime)$. One can then show that,
\begin{widetext}
\begin{align}
    g[\bm V](\bm e^\prime,\bm e_{\rm anc}^\prime)&\propto\exp\left[-\frac{1}{2}(\bm e^\prime\oplus\bm e^\prime_{\rm anc})^\top\bm V^{-1}(\bm e^\prime\oplus\bm e^\prime_{\rm anc})\right] \nonumber \\
    &=\exp\left[-\left(\frac{2G-1}{2\sigma^2}\right)\left(\bm e^\prime-\bm F\bm e_{\rm anc}^\prime\right)^2\right]
    \times \exp\left[-\frac{1}{2\sigma^2}\bm e_{\rm anc}^{\prime\,\top}\bm\Sigma_{\rm anc}^{-1}\bm e_{\rm anc}^\prime\right],\label{eq:g_expanded}
\end{align}
\end{widetext}
where, for convenience, we have defined
\begin{align}
\bm\Sigma_{\rm anc}\coloneqq\sigma^2\bm B^\top[(2G-1)\bm I_2\oplus \bm I_{2(K-1)}]\bm B.
\end{align} 
From here, we see that, conditioned on the `Gaussian estimator', $\bm F\bm e_{\rm anc}^\prime$, the data noise in each quadrature has been reduced to $\sigma^2/(2G-1)$. This interpretation is a bit premature as such does account for errors measurement stage, i.e. lattice effects from the ancillary GKP.

The output noise on the data, after the O2O layer of error correction, can be obtained from the error PDF, 
\begin{equation}\label{eq:Pe_prelim}
    P(\bm\varepsilon)=\int\dd\bm e\,\dd\bm{e_{\rm anc}}\,g[\bm V](\bm e^\prime,\bm e_{\rm anc}^\prime)\times\delta\left(\bm\varepsilon -\left(\bm e^\prime-\tilde{\bm e}^\prime\right)\right),
\end{equation}
where, recall, $\tilde{\bm e}^\prime=\bm F\bm\Omega^\top\bm s_{\rm anc}$ is the estimator informed by ancillary GKP syndromes, $\bm s_{\rm anc}=\bm \Omega\bm e_{\rm anc}^\prime\mod{\ell}$. Next, we write out the Dirac-delta distribution explicitly, expanding out the syndrome and accounting for modular effects in $\bm s_{\rm anc}$. First, for brevity, define the finite intervals $\mathscr{I}_{n_i}(\ell)\coloneqq[(n_i-1/2)\ell,(n_i+1/2)\ell]$ of characteristic size $\ell$, indexed by integers $n_i\in\mathbb{Z}$. Let $\mathscr{I}_{\bm n}(\ell)\coloneqq\prod_{i=1}^{2K}\mathscr{I}_{n_i}(\ell)$ define the region formed by their product. Then,
\begin{multline}
    \delta\left(\bm\varepsilon -\bm e^\prime+\tilde{\bm e}^\prime\right)=
    \sum_{\bm n\in\mathbb{Z}^{2K}}\delta\left(\bm\varepsilon-\bm e^\prime+\bm F\bm e_{\rm anc}^\prime-\ell\bm F\bm\Omega^\top \bm n\right) \\
    \times \mathcal{I}\left[\bm e^\prime_{\rm anc}\in\mathscr{I}_{\bm n}(\ell)\right],
\end{multline}
where $\mathcal{I}(\bm x\in\mathcal{X})=1$ if $\bm x\in\mathcal{X}$ and zero otherwise. The integers, $\bm n$, comes from the estimation procedure at the O2O layer, where displacements are measured from the nearest lattice points of the ancillary square GKP~\cite{noh2020o2o}. Substituting this expression into Eq.~\eqref{eq:Pe_prelim}, we find,
\begin{equation}\label{eq:Perror}
    P(\bm\varepsilon)=\sum_{\bm n\in\mathbb{Z}^{2K}}f(\bm n)\times g\left[\left(\frac{\sigma^2}{2G-1}\right)\bm I_2\right](\bm\varepsilon-\bm\mu(\bm n))
\end{equation}
where
\begin{equation}\label{eq:mun_dtms}
    \bm\mu(\bm n)\coloneqq \ell\bm F\bm\Omega^\top \bm n=\frac{\ell C_G}{\sqrt{K}}
    \begin{pmatrix}
        \sum_{i=\text{even}}n_i \\[.5em]
        \sum_{i=\text{odd}}n_i 
    \end{pmatrix},
\end{equation}
is a discretized mean, corresponding to Eq.~\eqref{eq:dtms_mu} in the main text, and we have defined the discrete probability distribution,
\begin{equation}\label{eq:fn_general}
    f(\bm n)\coloneqq \int_{\bm x\in\mathscr{I}_{\bm n}(\ell)}\dd{\bm x}g[\bm\Sigma_{\rm anc}](\bm x),
\end{equation}
where $C_G$ is defined in Eq.~\eqref{c_define}.
For instance, given a single GKP ancilla ($K=1$), $f(0,0)$ is the likelihood that an ancillary displacement lies within the square region $\mathscr{I}_{0,0}(\ell)=[-\ell/2,\ell/2]\times[-\ell/2,\ell/2]$ centered about the origin.


\subsubsection{Output error: Oscillators-to-oscillators}\label{appssec:o2o}

We now estimate the output error of a dtms-O2O code using similar tricks as in the original work of Ref.~\cite{noh2020o2o}.

In our current formulation, the $q$ and $p$ quadratures are uncorrelated and on equal footing. Thus, it suffices to focus on the noise in one quadrature, say the $q$ quadrature. We denote the error PDF for the $q$ quadrature as the marginal, $P(\varepsilon_q)=\int\dd{\varepsilon_p}P(\bm\varepsilon)$ where $\bm\varepsilon=\varepsilon_q\oplus\varepsilon_p$, such that $\sigma_{\rm out}^2=\int\dd{\varepsilon_q}P(\varepsilon_q)\varepsilon_q^2$.

For brevity, denote the $q$ component of the discretized mean, $\bm\mu(\bm n)$, as $\mu_q(\bm n)$ [first row of Eq.~\eqref{eq:mun_dtms}]. Then, given the error PDF in Eq.~\eqref{eq:Perror}, we can express the output variance of the dtms-O2O code formally as,
\begin{equation}
    \sigma_{\rm out}^2=\frac{\sigma^2}{2G-1}+\sum_{\bm n\in\mathbb{Z}^{2K}}f(\bm n)\mu_q^2(\bm n),
    \label{sigma_dmts_full_app}
\end{equation}
where the first term can be obtained entirely from the Gaussian approximation while the second term encodes lattice effects. This relation obeys the variance bound (i.e., no-threshold result) for oscillators-to-oscillators codes~\cite{hanggli2022NoThresholdGKP,wu2023optimal}.

We now make some approximations to obtain an asymptotic expression for the output error, which should be valid in the low-noise regime, $\sigma\ll 1$. In particular, we suppose that $G\sigma^2/K\lesssim \ell^2$. For low noises, we ignore $|n_i|\geq 2$ terms of the summation in Eq.~\eqref{sigma_dmts_full_app}, thus focusing on $n_i=\pm 1$ type of terms. Observe that, for low noise and high gain ($G\sim K\ell^2/\sigma^2$), the noises on the GKP ancillary modes will be highly correlated, due to uniformly distributing the single amplified noise on the first ancilla to all $K$ ancillary modes. This means the term $f(1, 1, \dots, 1)$ will dominate the summation. [The final expression depends weakly on this assumption. It can nevertheless be validated by observing the form of $\bm\Sigma_{\rm anc}$.] For this set of indices, we have $\mu_q(1,\dots,1)=C_G\sqrt{K}$ [see Eq.~\eqref{c_define}]. Then, using $f(-\bm n)=f(\bm n)$ and $f(1,\dots,1)\leq f(1)$, where $f(1)=\sum_{\bm n^\prime\in\mathbb{Z}^{2(K-1)}} f(1,\bm n^\prime)$ is the marginal, we can expand the output error as,
\begin{align}
    \sigma^2_{\rm out}\approx \frac{\sigma^2}{2G-1} + \frac{16\pi G(G-1)K}{(2G-1)^2}f(1).
\end{align}
The goal now is to estimate $f(1)$, which amounts to taking the marginal over $f(\bm n)$ in Eq.~\eqref{eq:fn_general}. It can be verified that the single-quadrature ancilla variance is $\sigma^2_{\rm anc}=\sigma^2(2(G-1)/K + 1)$, and thus,
\begin{equation}\label{eq:f1_marginal}
    f(1)=\int_{\ell/2}^{3\ell/2}g[\sigma^2_{\rm anc}](x) \approx \frac{1}{2}{\rm erfc}\left(\sqrt{\frac{\ell^2}{8\sigma_{\rm anc}^2}}\right),
\end{equation}
where the approximation is from taking the upper limit to infinity. Expanding everything out in full, we find
\begin{widetext}
\begin{equation}\label{eq:sig_out_almost}
    \sigma_{\rm out}^2\approx \frac{\sigma^2}{2G-1} +\frac{8\pi G(G-1)K}{(2G-1)^2}{\rm erfc}\left(\frac{1}{2}\sqrt{\frac{\pi}{\sigma^2\left(\frac{2(G-1)}{K}+1\right)}}\right).
\end{equation}
\end{widetext}
Using the approximation ${\rm erfc}(x)\approx e^{-x^2}/(\sqrt{\pi}x)$, it can be shown \QZ{ by iteration} that the optimal gain, $G_{\rm O2O}$, that minimizes the output error is approximately given as,
\begin{equation}
    G_{\rm O2O}\approx \frac{\pi K}{8\sigma^2}\left\{\ln\left(\frac{\pi^{3/2}K^2}{2\sigma^4}\right)\right\}^{-1}.
\end{equation}
The first part minimizes the exponential in Eq.~\eqref{eq:sig_out_almost}, while the logarithmic term accounts for the pre-factor.

\subsubsection{Qudit layer decoding: Error rates}

We estimate single-qudit error rates assuming a square qudit as the data mode in the (CSS) dtms-GKP code. 

Following O2O-level decoding (first stage of the decoder), we perform closest point decoding on the square qudit. In other words, we measure $q$ and $p$ stabilizers of the square qudit code to estimate the displacement and map back to the nearest lattice point. Since the code is CSS in our current formulation, we can focus on one type of error, say an $X$ error, which is governed by the marginal PDF, $P(\varepsilon_q)$. An error occurs if the displacement, $\varepsilon_q$, falls within an error region, $\mathscr{E}_d(m)\coloneqq \ell/\sqrt{d}[md+1/2,(m+1)d-1/2]$, where $d$ is the qudit dimension and $m\in\mathbb{Z}$. That is, if $\varepsilon_q\in\mathscr{E}_d(m)$, closest point decoding will map us back to an incorrect lattice point, inducing a logical error. Using the form of the error PDF provided in Eq.~\eqref{eq:Perror}, the logical error rate can be formally written as,
\begin{align}
\Pr[X]&=\sum_{m\in\mathbb{Z}}\int_{\mathscr{E}_d(m)}\dd{\varepsilon_q} P(\varepsilon_q) \nonumber \\
    &=\sum_{\bm n\in\mathbb{Z}^{2K}}f(\bm n)\Pr[X|\bm n],
\end{align}
where we have defined the conditional error rate, 
\begin{widetext}
\begin{align}
    \Pr[X|\bm n]&\coloneqq \sum_{m\in\mathbb{Z}} \int_{\mathscr{E}_d(m)}\dd{\varepsilon_q}g\left[\frac{\sigma^2}{2G-1}\right]\left(\varepsilon_q-\mu_q(\bm n)\right)\nonumber\\
    &= \frac{1}{2}\sum_{m\in\mathbb{Z}}\left({\rm erf}\left[\frac{\frac{\ell}{\sqrt{d}}((m+1)d-\frac{1}{2})-\mu_q(\bm n)}{\sqrt{2\sigma^2/(2G-1)}}\right] - {\rm erf}\left[\frac{\frac{\ell}{\sqrt{d}}(md+\frac{1}{2})-\mu_q(\bm n)}{\sqrt{2\sigma^2/(2G-1)}}\right]\right).
\end{align}
\end{widetext}
At $G=1$, these expressions result in error rates for the square qudit code~\cite{gkp2001}, $\eval{\Pr[X]}_{G=1}\approx {\rm erfc}[\sqrt{\ell^2/(8d\sigma^2)}]$. We can then imagine starting from $G=1$ and slowly increasing the gain, thereby decreasing the error rate. For $G$ close to 1, effects from the O2O decoding layer are negligible and $\Pr[X]\approx {\rm erfc}[\sqrt{\ell^2(2G-1)/(8d\sigma^2)}]$. On the other hand, for $G \gtrsim 1$, errors are dominated by the O2O-layer decoding and thus determined by $f(\bm n)$. In particular,
\begin{equation}
    \eval{\Pr[X]}_{G\gtrsim 1} \approx \sum_{\bm n\neq (0,\dots,0)}f(\bm n).
\end{equation}
Assuming the gain is larger than 1, but not too large, specifically $G\lesssim K$, then only first order terms in the sum are necessary, i.e.
\begin{align}
    \sum_{\bm n\neq\bm 0}f(\bm n).
    &\approx 2\times\left(f(1,0,\dots,0)+\dots + f(0,0,\dots,1)\right) \nonumber\\
    &\approx 2K \times f(1),
\end{align}
where the factor of 2 comes from $f(-\bm n)=f(\bm n)$ symmetry and $f(1)$ is the marginal, $f(1)=\sum_{\bm n^\prime} f(1, \bm n^\prime)$, given in Eq.~\eqref{eq:f1_marginal}. Expanding the error rate out in these different regimes, we have, 
\begin{equation}
    \Pr[X]\approx 
    \begin{cases}
        {\rm erfc}\left[\sqrt{\frac{\ell^2(2G-1)}{8d\sigma^2}}\right], & G\approx 1 \\[1.5em] 
        K\times{\rm erfc}\left[\sqrt{\frac{\ell^2}{8\sigma^2\left(\frac{2(G-1)}{K}+1\right)}}\right], & 1 \lesssim G \lesssim K
    \end{cases}.
\end{equation}
These cases are (approximately) balanced when the arguments of the exponentials are equal. This happens at the gain value,
\begin{equation}
    G(N;d)=1+\frac{\sqrt{(N-2)^2+4d(N-1)}-N}{4},
\end{equation}
where $N=K+1$. Setting this as the gain once and for all, we find,
\begin{align}
    \Pr[X]&\approx (K+1)\times {\rm erfc}\left[\sqrt{\frac{\ell^2\left(2G(N;d)-1\right)}{8d\sigma^2}}\right] \\ 
    &=N\times {\rm erfc}\left[\sqrt{\frac{\mathfrak{D}_{\rm eff}^2(N;d)}{8\sigma^2}}\right],
\end{align}
where we have defined the effective distance, $\mathfrak{D}_{\rm eff}(N;d)\coloneqq\ell\sqrt{(2G(N;d)-1)/d}$. These expressions agree with those provided in Section~\ref{ssec:ex2_qudit}.

Similar asymptotic analyses can be applied to dtms two-qubit codes, though we do not provide details here. Notably, the main difference for two-qubit codes is the existence of correlations between $X_1$ and $X_2$ ($Z_1$ and $Z_2$) errors, which must be taken into consideration. However, the error rate for a single-qubit error, say $X_1$, closely resembles the single-qudit case just discussed. Hence, similar arguments go through.

\section{Assorted Code Examples}\label{app:example_codes}

We present various GKP qubit codes, such as the tesseract code, the [[4,2,2]] code, and the [[5,1,3]] code, along with symplectic transformations that can be used to construct them. This gives some insight into relationships between different codes and demonstrates how qubit codes can be created via Gaussian operations on a square qubits and canonical GKP states. To streamline the presentation, we use the following notation for the initial generator matrix of a $[[N,k]]$ qubit code,
\begin{equation}
    \bm{M}_{(N,k)}\coloneqq \bm M_{\rm in} =\left(\sqrt{2}\bm{I}_{2k}\right) \oplus \bm{I}_{2(N-k)}.
\end{equation}

\subsection{Tesseract code}
The Tesseract code can be represented by the generator matrix~\cite{baptiste2022multiGKP} 
\begin{align}
    \bM_{\rm{tess}} = 2^{1/4}\begin{pmatrix}
        1 & 0 & 0 & 0\\
        0 & 1/\sqrt{2} & 0 & 1/\sqrt{2} \\
        0 & 0 & -1 & 0\\
        0 & 1/\sqrt{2} & 0 & -1/\sqrt{2}
    \end{pmatrix}.
\end{align}
It can also be constructed by acting with a two-mode Gaussian operation, $\bS_{\rm{tess}}$, on a square GKP qubit and a canonical GKP state, i.e. $\bm M^\prime_{\rm tess}=\bm S_{\rm tess}\bm M_{(2,1)}$, where
\begin{equation}
        \bS_{\rm{tess}} = \begin{pmatrix}
            2^{-1/4} & 0 & 2^{1/4} & 0\\
            0 & 0 & 0 & 2^{-1/4} \\
            -2^{-1/4} & 0 & 0 & 0\\
            0 & -2^{1/4} & 0 & 2^{-1/4}
        \end{pmatrix}.
\end{equation}
The generator matrices, $\bm M_{\rm tess}$ and $\bm M_{\rm tess}^\prime$, are related by a unimodular matrix,
\begin{equation}
    \bN_{\rm{tess}}=\begin{pmatrix}
            1& 0 & 1 & 0\\
            0&-1 & 0 & 1\\
            1& 0 & 0& 0\\
            0& 1& 0 & 0
        \end{pmatrix},
\end{equation}
such that $\bS_{\rm tess} \bm{M}_{(2,1)} =\bM_{\rm tess}\bN_{\rm {tess}}$. Furthermore, it can be shown that the Tesseract symplectic transformation, $\bm S_{\rm tess}$, can be decomposed into a two-mode squeezer followed by a beamsplitter. The relation can be expressed as $\bS_{\rm tess}=\bB_{\rm tess}\bS_{G_{\rm tess}}$, with $G_{\rm tess} = (\sqrt{2}+1)/2$ and
\begin{equation}
         \bB_{\rm tess} =\begin{pmatrix}
            \sqrt{\eta} \bI_2 & \sqrt{1-\eta} \bI_2 \\
            -\sqrt{1-\eta} \bI_2 & \sqrt{\eta} \bI_2
        \end{pmatrix},
\end{equation}
where $\eta = (2-\sqrt{2})/4$. Since the code distance (and error correction for iid AGN) does not change with the exterior beamsplitter, $\bB_{\rm tess}$, the Tesseract code is equivalent to the $N=2$ dtms qubit code in terms of code distance. In fact, the code distance is $2^{1/4} \sqrt{\pi}$ for both codes.

\subsection{[[4,2,2]] code}
From the [[4,2,2]] stabilizers $\langle X X X X , Z Z Z Z\rangle$, a generator matrix can be written~\cite{eisert2022lattice,baptiste2022multiGKP,brady2023GKPrvw}, 
\begin{equation}
    \bM_{[[4,2,2]]}=\frac{1}{\sqrt{2}}\begin{pmatrix}
        1 & 0 & 2 & 0 & 0 & 0 &0 &0\\
        0 & 1 & 0 & 2 & 0 & 0 &0 &0\\
        1 & 0 & 0 & 0 & 2 & 0 &0 &0\\
        0 & 1 & 0 & 0 & 0 & 2 &0 &0\\
        1 & 0 & 0 & 0 & 0 & 0 &2 &0\\
        0 & 1 & 0 & 0 & 0 & 0 &0 &2\\    
        1 & 0 & 0 & 0 & 0 & 0 &0 &0\\
        0 & 1 & 0 & 0 & 0 & 0 &0 &0\\           
    \end{pmatrix}.
\end{equation}
The [[4,2,2]] code can also be constructed by a symplectic transformation, $\bS_{[[4,2,2]]}$, such that $\bS_{[[4,2,2]]}\bm{M}_{(4,2)}=\bM_{[[4,2,2]]}\bN_{[[4,2,2]]}$, where
\begin{subequations}
    \begin{align}
        & \bS_{[[4,2,2]]} =\begin{pmatrix}
            1 & 0 & 0 & -1 & \frac{1}{\sqrt{2}} & 0 & \sqrt{2} & 0 \\
            0 & 0 & 0 & 0 & 0 & 0 & 0 & \frac{1}{\sqrt{2}}\\
            1 & 0 & 0 & 0 & \frac{1}{\sqrt{2}} & 0 & 0 & 0\\
            0 & 1 & 0 & 0 & 0 & 0 & 0 & -\frac{1}{\sqrt{2}}\\
            0 & 0 & 0 & 0 & \frac{1}{\sqrt{2}} & 0 & 0 & 0\\
            0 & -1 & -1 & 0 & 0 & \sqrt{2} & 0 & \frac{1}{\sqrt{2}}\\
            0 & 0 & 0 & -1 & \frac{1}{\sqrt{2}} & 0 & 0 & 0\\
            0 & 0 & 1 & 0 & 0 & 0 & 0 & -\frac{1}{\sqrt{2}}\\
        \end{pmatrix},\\
        & \bN_{[[4,2,2]]} =\begin{pmatrix}
            0 & 0 & 0 & -2 & 1 & 0 & 0 & 0\\
            0 & 0 & 2 & 0 & 0 & 0 & 0 & -1\\
            1 & 0 & 0 & 0 & 0 & 0 & 1 & 0\\
            0 & 0 & -1 & 0 & 0 & 0 & 0 & 1\\
            1 & 0 & 0 & 1 & 0 & 0 & 0 & 0\\
            0 & 1 & -1 & 0 & 0 & 0 & 0 & 0\\
            0 & 0 & 0 & 1 & 0 & 0 & 0 & 0\\
            0 & -1 & -2 & 0 & 0 & 1 & 0 & 1\\
        \end{pmatrix}. 
    \end{align}
\end{subequations}
The code distance is $\sqrt{2\pi}$. Note that the [[4,1,2]] code is a sub-code of the [[4,2,2]] code with the same GKP code distance.


\subsection{[[5,1,3]] code}
The stabilizer generators of the $[[5,1,3]]$ code are $\langle IXZZX,XZZXI,XIXZZ,ZXIXZ\rangle$, thus providing a lattice representation, 
\begin{equation}
    \bM_{[[5,1,3]]} = \frac{1}{\sqrt{2}}\begin{pmatrix}
        0 & 1 & 1 & 0 & 2 & 0 & 0 & 0 & 0 & 0\\
        0 & 0 & 0 & 1 & 0 & 2 & 0 & 0 & 0 & 0\\
        1 & 0 & 0 & 1 & 0 & 0 & 2 & 0 & 0 & 0\\
        0 & 1 & 0 & 1 & 0 & 0 & 0 & 2 & 0 & 0\\
        0 & 0 & 1 & 0 & 0 & 0 & 0 & 0 & 2 & 0\\
        1 & 1 & 0 & 0 & 0 & 0 & 0 & 0 & 0 & 2\\
        0 & 1 & 0 & 1 & 0 & 0 & 0 & 0 & 0 & 0\\
        1 & 0 & 1 & 0 & 0 & 0 & 0 & 0 & 0 & 0\\
        1 & 0 & 0 & 0 & 0 & 0 & 0 & 0 & 0 & 0\\
        0 & 0 & 1 & 1 & 0 & 0 & 0 & 0 & 0 & 0\\
    \end{pmatrix}.
\end{equation}
One can also obtain the [[5,1,3]] code from a symplectic transformation, $\bS_{[[5,1,3]]}$, such that  ${\bS_{[[5,1,3]]}\bM_{(5,1)}=\bM_{[[5,1,3]]}\bN_{[[5,1,3]]}}$. The symplectic matrix, $\bS_{[[5,1,3]]}$, and unimodular matrix, $\bN_{[[5,1,3]]}$, are given explicitly as,
\begin{widetext}
\begin{equation}
        \bS_{[[5,1,3]]} = \begin{pmatrix}
            0 & -1 & \frac{1}{\sqrt{2}} & 0 & \sqrt{2} & 0 & 0 & \frac{1}{\sqrt{2}} & 0 & 0 \\
            0 & 0 & 0 & 0 & 0 & \frac{1}{\sqrt{2}} & 0 & 0 & 0 & 0 \\
            0 & 0 & 0 & 0 & 0 & 0 & 0 & 0 & \frac{1}{\sqrt{2}} & 0 \\
            0 & 0 & 0 & 0 & 0 & 0 & 0 & \frac{1}{\sqrt{2}} & 0 & \sqrt{2} \\
            0 & 0 & \frac{1}{\sqrt{2}} & 0 & 0 & 0 & 0 & 0 & 0 & 0 \\
            -1 & 0 & 0 & \sqrt{2} & 0 & -\frac{1}{\sqrt{2}} & -\sqrt{2} & \frac{1}{\sqrt{2}} & \frac{1}{\sqrt{2}} & 0 \\
            -1 & -1 & 0 & 0 & 0 & \frac{1}{\sqrt{2}} & 0 & \frac{1}{\sqrt{2}} & 0 & 0 \\
            0 & 0 & \frac{1}{\sqrt{2}} & 0 & 0 & -\frac{1}{\sqrt{2}} & -\sqrt{2} & 0 & \frac{1}{\sqrt{2}} & 0 \\
            0 & 1 & 0 & 0 & 0 & -\frac{1}{\sqrt{2}} & -\sqrt{2} & 0 & \frac{1}{\sqrt{2}} & 0 \\
            -1 & -1 & \frac{1}{\sqrt{2}} & 0 & 0 & \frac{1}{\sqrt{2}} & 0 & 0 & 0 & 0 \\
        \end{pmatrix},
\end{equation}
\end{widetext}
and
\begin{equation}
\bN_{[[5,1,3]]} = \begin{pmatrix}
            0 & 2 & 0 & 0 & 0 & -1 & -2 & 0 & 1 & 0\\
            0 & -2 & 0 & 0 & 0 & 0 & 0 & 1 & 0 & 0\\
            0 & -2 & 1 & 0 & 0 & 0 & 0 & 0 & 0 & 0\\
            -2 & 0 & 0 & 0 & 0 & 1 & 0 & 0 & 0 & 0\\
            0 & 1 & 0 & 0 & 1 & 0 & 0 & 0 & 0 & 0\\
            1 & 0 & 0 & 0 & 0 & 0 & 0 & 0 & 0 & 0\\
            1 & -1 & 0 & 0 & 0 & 0 & 1 & 0 & 0 & 0\\
            0 & 1 & 0 & 0 & 0 & 0 & 1 & 0 & 0 & 1\\
            0 & 1 & 0 & 0 & 0 & 0 & 1 & 0 & 0 & 0\\
            -1 & 0 & 0 & 1 & 0 & 0 & 0 & 0 & 0 & 0\\
        \end{pmatrix}.
\end{equation}

The code distance of the [[5,1,3]] code is $\sqrt{3}\mathfrak{D}_{\square}=\sqrt{3\pi}$.




\end{document}